\documentclass{elsart}

\usepackage{graphicx}
\usepackage{epsf}
\usepackage{epsfig}
 
\usepackage{amssymb}

\usepackage{lineno}

\begin{document}



\begin{frontmatter}
\begin{boldmath}
\title{Energy Calibration of $b$-Quark Jets
             with $Z \to b\bar{b}$ Decays
             at the Tevatron Collider}
\end{boldmath}

\author[Padova]{J. Donini\corauthref{cor}},
\corauth[cor]{Corresponding author. Address: Dipartimento di Fisica ``Galileo Galilei'', via F. Marzolo, 8 - 35131 Padova, Italy. Telephone number: +39 338 59 49 675. Fax number: +39 49 827 7237.}
\ead{donini@pd.infn.it}
\author[Padova]{T. Dorigo},
\author[Rockefeller]{K. Hatakeyama},
\author[Chicago]{S. Kwang},
\author[Penn]{C. Neu},
\author[Chicago]{M. Shochet},
\author[Tsukuba]{T. Tomura},
\author[Padova]{M. Tosi},
\author[Penn]{D. Whiteson}

\address[Padova]{University of Padova and INFN, I-35131 Padova, Italy}
\address[Rockefeller]{The Rockefeller University, New York, New York, 10021, USA}
\address[Chicago]{Enrico Fermi Institute, University of Chicago, Chicago, IL 60637, USA}
\address[Penn]{University of Pennsylvania, Philadelphia, PA 19104, USA}
\address[Tsukuba]{University of Tsukuba, Tsukuba, Ibaraki 305-8571, Japan}

\begin{abstract}
The energy measurement of jets produced by $b$-quarks at
hadron colliders suffers from biases due to the peculiarities of the
hadronization and decay of the originating $B$ hadron. 
The impact of these effects can be estimated by reconstructing the
mass of $Z$ boson decays into pairs of $b$-quark jets.
From a sample of $584 \ {\rm pb^{-1}}$ of data collected by the CDF
experiment in $1.96 \ {\rm TeV}$ proton-antiproton collisions at the
Tevatron collider, we show how the $Z$ signal can be identified and
measured. 
Using the reconstructed mass of $Z$ candidates we determine a jet energy scale factor for
$b$-quark jets with a precision better than $2\%$. 
This measurement allows a reduction of one of the dominant
source of uncertainty in analyses based on high transverse momentum $b$-quark jets.
We also determine, as a cross-check of our analysis, the $Z$ boson cross
section in hadronic collisions using the $b \bar b$ final state as
$\sigma_Z \times B(Z \to b\bar b) = 1578^{+636}_{-410}$ pb. 
\end{abstract}

\begin{keyword}
$b$-jet energy scale \sep $Z$ boson \sep Collider Detector at Fermilab  \sep CDF \sep Fermilab \sep High-energy physics

\PACS 14.70.Hp \sep 14.65.Fy \sep 14.65.Ha \sep 13.87.Fh
\end{keyword}

\end{frontmatter}

%

\section {Introduction}\label{s:introduction}

\noindent Since their discovery in 1983~\cite{wzdisc},
$W$ and $Z$ bosons have been studied
at hadron colliders predominantly using their leptonic decays.
In fact, the hadronic decays of these particles are generally so difficult
to separate from the large background arising from generic jet pairs produced by
quantum chromodynamics (QCD) interactions that, after the extraction
of a mass bump in the dijet mass distribution
by the UA2 collaboration from 630 GeV data at the $Sp\bar p S$
in 1987~\cite{wzjjbump}, little more has emerged.
At the Fermilab Tevatron the direct observation of hadronic decays
of vector bosons is more difficult. With respect to the
$Sp\bar{p}S$, the Tevatron's higher center-of-mass energy is a
disadvantage because, although the signal cross section is four
times larger, the irreducible background from QCD processes yielding
jet pairs increases by over an order of magnitude.

During Run I at the Tevatron (1992-96) hadronic $W$ decays were
successfully used by the CDF and D\O\ experiments in the discovery
and measurement of the top quark both in the single lepton and fully
hadronic final states.
The larger Run II (started in 2001) data sample made it possible to exploit the
hadronic decay of $W$ bosons in top events for a direct calibration
of the energy measurement of light-quark jets in the reconstruction
of the $t\bar{t}$ decay. That technique provides a
significant reduction of the systematic uncertainty arising from the
jet energy measurement~\cite{2dtmass}.
Tevatron experiments can indeed reach an accuracy close to 1
GeV/$c^{2}$ on the top quark mass in Run II by reducing to about 1\% the uncertainty on the jet energy scale (JES),  
a factor which measures the discrepancy between the effect of detector response and energy
corrections in real and simulated hadronic jets.

For the $Z$ boson, which is not produced in top decays,
the observation in hadronic final states is more complicated. Only
the decay to identified $b$-quark pairs is observable because 
the enormous $gg \to gg$ background becomes significantly reduced. Indeed, a
small signal of $Z \to b\bar{b}$ decays was extracted by CDF in Run
I data exploiting the semileptonic decay of $b$ quarks by triggering
on muons with low transverse momentum~\cite{zbbr1}. The
signal was too small to extract information on the accuracy of the
$b$-jet energy measurement, but it spurred the development of a
dedicated trigger in Run II. A large signal of $Z \to b\bar{b}$
decays free from selection biases allows a precise measurement of
the energy scale of $b$-quark jets and provides a determination
of the $b$-jet energy resolution. The reduction of the uncertainty
in the $b$-jet energy scale ($b$-JES), that measures the ratio between
the calorimeter response in real and simulated $b$-quark jets, can help
all precision measurements of the top quark mass, while a
determination of the $b$-jet energy resolution is important for the
search for a low-mass Higgs boson. The signal allows a direct tuning
of algorithms that seek to increase the resolution of the $b$-jet
energy measurement. These algorithms are a critical ingredient for
the potential observation of the Higgs boson at the Tevatron if $M_H<135$
GeV/$c^{2}$.

The largest contribution to the total uncertainty in the
top quark mass determination at the Tevatron originates from the
knowledge of the jet energy scale.
The JES can be determined from studies of detector response to single particles and
a detailed modeling of their $P_T$ spectrum in jets~\cite{cdfjesNIM}. An
estimate of the accuracy of the resulting calibration can then be
obtained with events containing a single photon recoiling against a
hadronic jet, but the method is limited by systematic uncertainties
arising from the modeling of the production process. 
A check of the light-quark calibration, which is essentially statistics-limited, comes instead from the
measurement of $W \to q \bar{q}'$ decays in $t\bar{t}$ events, as mentioned earlier.

When dealing with $b$-quark jets one has to cope with
their unique fragmentation and decay properties, that result in uncertainties in the jets energy response.
The $B$ hadron resulting from $b$-quark fragmentation carries a larger fraction of the parent quark momentum than hadrons originating from light quark fragmentation. 
Uncertainties in the jet energy response arise from the imperfect knowledge of the fragmentation properties of $b$-quarks.
Moreover $b$-jets have a different response on average than light-quark and gluon jets because of the large semi-leptonic decay fraction of $B$ hadrons. 
The imperfect knowledge of the decay properties of $b$-quarks yield an additional uncertainty on the jet energy response.
Those effects have to be accurately modeled if one is to apply a
generic JES factor, derived from light-quark and gluon jets, to the
two $b$-jets emitted in $t\bar{t}$ decays.

Due to the small cross section of production processes yielding
events with a high-energy photon recoiling against a $b$-quark jet,
a measurement of the $b$-JES with
transverse balancing techniques~\cite{cdfjesNIM} is quite difficult,
and the Tevatron experiments have so far been unable to exploit them.
In this paper, however, we demonstrate the feasibility of extracting the
$b$-JES from the reconstructed $Z \to b\bar{b}$ signal. We first discuss in detail the data sample we use and
its collection and reconstruction; then we describe the method by
which we model the spectrum of the huge background from
QCD events and our fitting technique. We then evaluate the
systematic uncertainties affecting the signal extraction.
Finally, we present our measurement of the $b$-JES factor, which achieves an uncertainty
better than $2\%$ using an integrated luminosity of
$584 \ {\rm pb^{-1}}$ of CDF Run II data, and we provide for the
first time, as a cross check of our analysis, a measurement of the $Z$ boson cross section from its $b \bar
b$ final state.

\section{The detector and the datasets}\label{s:data}

In this section we provide a short description of the experimental
facility and discuss the method by which the sample of data used for
our analysis is collected. We also describe the software
reconstruction of the event characteristics that are most relevant
to the measurement of $Z$ boson decays to $b$-quark pairs.

\subsection {The CDF detector}

CDF is a magnetic spectrometer designed to detect and measure, over
a wide range of rapidity, charged and neutral particles produced by
1.96 TeV proton-antiproton collisions delivered by the Tevatron
collider. The detector is described in detail
elsewhere~\cite{cdfdet}; in this section we only describe those
components most relevant to the $Z \to b\bar{b}$ measurement.

A seven-layer silicon vertex detector (SVX) is located immediately
outside of the beam pipe. Its microstrip sensors measure with a precision of about 
ten micrometers the position where charged particles cross each layer; that information allows the
discrimination of tracks originated from the decay of long-lived
particles, as explained in section \ref{s:evtrec}. 
Outside the silicon detector, a large cylindrical multilayer drift chamber, the Central Outer Tracker (COT),
measures track momenta within the pseudorapidity~\cite{cdfsystem} interval $|\eta|<1.0$
from the curvature of their helices in the $1.4$ T axial magnetic field. 
Electromagnetic and hadronic sampling calorimeters, arranged in a 
projective-tower geometry, surround the tracking systems and measure the 
energy and direction of electrons, photons, and jets in the range 
$|\eta|<3.6$. Muon systems outside the calorimeters allow the 
reconstruction of track segments for penetrating particles within $|\eta|<1.5$. 
The beam luminosity is determined using 
gas Cherenkov counters surrounding the beam pipe, which measure the 
average number of inelastic $p\bar{p}$ collisions per bunch crossing.

Data acquisition is initiated by a three-level trigger system~\cite{cdftrigger}. Level 1 and
level 2 consist of dedicated hardware modules, while level 3 runs speed-optimized
reconstruction algorithms on a farm of commercial processors.
Of particular relevance to our analysis are the hardware devices
reconstructing charged particle tracks in the level 1 (XFT) and level 2
(SVT) trigger systems. XFT~\cite{xft}, the extremely fast tracker,
uses hits in the COT and a fast hardware architecture to measure
transverse momentum and azimuthal angle of charged tracks. 
SVT~\cite{svt}, the Silicon Vertex Trigger, is a highly parallel
system which allows the measurement of the impact parameter of tracks
using SVX information, with precision (35 $\mu$m for 2 GeV/$c$ tracks) similar
to that obtained offline. 

\subsection{The $Z \to b\bar{b}$ trigger }\label{s:trigger}

A trigger (called Z\_BB), selecting events with low transverse energy jet pairs 
and containing tracks with significant impact parameter,
was implemented in Run II
as a mean of acquiring a large sample of $Z$
decays to pairs of $b$-jets.
The trigger design was based
on the ability of the SVT to identify and measure tracks with a
significant value of $d_0$, the impact parameter with respect to the beam position in
the transverse plane.

During its operation,
the trigger underwent a few small modifications to maintain a manageable
trigger rate as the Tevatron instantaneous luminosity increased.
Because even small trigger modifications may affect the shape of the
background dijet mass distribution in a way which is very difficult
to model with the necessary accuracy, this measurement focuses on 
approximately half of the total data collected in the years 2001-2006, corresponding to a period of
data taking when the trigger did not undergo significant changes.

The version of the Z\_BB trigger that collected our data requires two
XFT tracks and a localized calorimeter energy deposit at level 1,
two low-$E_T$ calorimeter clusters~\cite{cdfdet} and two displaced SVT tracks at
level 2, and two reconstructed jets and two large impact parameter
tracks at level 3. Specifically, the following selection is applied
to each event:

\begin{itemize}
  \item level 1 requires one central ($|\eta|<1.1$) calorimeter tower with transverse energy 
	$E_T>5$ GeV,
      	plus two XFT tracks with transverse momentum above 5.5 and 2.5 GeV$/c$, respectively;
  \item level 2 vetoes events containing a calorimeter cluster of $E_T>5$ GeV 
	in the pseudorapidity range
    	$1.1<|\eta|<3.6$. Events are required to have two central $E_T>3$ GeV clusters 
      	plus two SVT tracks with $P_T>2$ GeV$/c$ and impact parameter 
        $d_0$ greater than 160 ${\rm \mu m}$.
  \item level 3 requires two central jets 
        having $E_T>10$ GeV.
        The event must also contain
        two tracks with $P_T>2$ GeV$/c$ and $d_0>160$ ${\rm \mu m}$;
        alternatively, track pairs with looser cuts ($P_T>1.5$ GeV$/c$, $d_0>130$ ${\rm \mu
        m}$) are accepted if the impact parameter is more than three times larger than its
        estimated measurement error.
\end{itemize}

\noindent A dynamic prescaling  was applied in the level 2 trigger
during a portion of the period of activity of the above trigger.
Dynamic prescaling automatically rejects a variable fraction of the
data passed by the trigger; the fraction depends on the
instantaneous luminosity of the colliding beams, and it is tuned to
keep the output rate of the trigger system within the storage
capabilities of the data acquisition system. The main effect of the
prescaling factor is to reduce the effective integrated luminosity
of the data; for the Z\_BB trigger the maximum reduction is a
factor of 10 at the highest luminosity. 

After a selection of runs tagged as ``good'' by the data quality
monitor, the dataset contains a total of 39 million events.
This corresponds to an integrated luminosity $L=584$ pb$^{-1}$ for
the Z\_BB trigger and the run range considered in our analysis.
The total uncertainty on the integrated luminosity is $5.9\%$.

\subsection{Monte Carlo samples}\label{s:mc}

Version 6.216 of the Pythia Monte Carlo (MC) program~\cite{pyt6216} was
used to generate 7.4 million $Z\to b\bar{b}$ events with
minimum-bias interactions superimposed according to the instantaneous luminosity
of the real data. The $Z$ boson sample was generated using CTEQ5L~\cite{CTEQ} parton density function (PDF) set.
The Pythia tune A parameter set~\cite{RFieldTA} was used to model the underlying event. 
Simulation of $B$ hadron decays was performed using the EvtGen~\cite{EvtGen} event generator.
Event reconstruction was performed with the same offline algorithms used for real data. 
Trigger requirements were emulated using trigger primitives simulation (level 1 and level 2) and offline variables (level 3).

In addition to the $Z\to b\bar{b}$ dataset, smaller samples of $Z\to
c\bar{c}$ and $W\to c\bar{s}$ were also generated and reconstructed
with the same recipe outlined above, to study the contamination of
such processes in our data sample (see Sec.~\ref{s:rescont}). Moreover,
several QCD Pythia Monte Carlo samples were used for additional
studies on sample composition and background modeling. Finally,
$Z\to e^+e^-$ and $Z\to \mu^+\mu^-$ MC samples, also
generated with Pythia, were used for studies of the systematic
uncertainty on signal acceptance due to initial state QCD radiation.
These studies are described in section \ref{s:effsig}.


\subsection {Event reconstruction}\label{s:evtrec}

Hadronic jets are reconstructed from calorimeter tower information
using an iterative jet cone clustering algorithm,
JetClu~\cite{jetclu}, with the cone radius
$R=\sqrt{\Delta\eta^2+\Delta\phi^2}=0.7$ units in the azimuth-pseudorapidity space.
Jet $E_T$ is computed by summing the energy deposited in
calorimeter towers within the cone multiplied by $\sin\theta$,
where $\theta$ is the polar angle of the $E_T$-weighted centroid
of the clustered tower.

The standard CDF jet energy correction package~\cite{cdfjesNIM}
determines the most probable energy of the parton that produced the
jet by applying to the ``raw'' jet energy (which is labeled level 0)
several factors in series, to account for detector non-uniformities
(level 1), multiple $p\bar{p}$ interactions (level 4),
calorimeter stability and non-linear response, fragmentation model and Monte Carlo tuning (level 5)
and other effects such as energy from the underlying event included
in the jet cone (level 6) and energy lost out of the clustering cone (level 7). 

Usually, CDF analyses seeking the reconstruction of massive object decays use jet energies corrected up to level 5 to 
reconstruct the kinematics of the final state, and apply separately custom corrections for those effects that are process-dependent;
that is the case, for instance, of top quark mass measurements. We made the same choice of correction level in our analysis in order to 
measure a $b$-jet energy scale factor that can be applied to other analyses based on high $P_T$ $b$-quark jets (provided that they use the same
$R=0.7$ jet cone definition).
From now on, ``corrected'' jet energies will imply jet energies corrected to level 5, while ``uncorrected'' jet energies will stand for 
raw (level 0) measured jet energies.


For the purpose of maximizing the heavy flavor content in the dijet
sample, we use the SecVtx $b$-tagging algorithm to search for secondary vertices. 
Indeed, the long lifetime and high mass of $B$ hadrons allow their
decays to be well displaced from the primary $p \bar p$ interaction
point, and thus form a secondary vertex.
The algorithm is described in detail elsewhere~\cite{cdfdet}. In short, SecVtx
searches for secondary vertices in jets with uncorrected $E_T>15$
GeV and pseudorapidity in the range $|\eta|<2.0$. 
The algorithm first selects charged tracks within the jet cone that have
been measured in the silicon detector with sufficiently good
position accuracy and that have a large significance of their
impact parameter with respect to the interaction point. The algorithm then uses these tracks in a
recursive procedure to reconstruct a common point of origin for at
least three of them. If the reconstruction fails, tighter
requirements are imposed on the tracks, and a fit accepting two-tracks vertices
is attempted. Reconstructed vertices are rejected if their transverse
distance from the interaction point corresponds to the location of
material of the innermost silicon layer (1.2 cm $< r <$ 1.5 cm) or if it is greater 
than 2.5 cm.
If a good vertex is found, several quantities are computed
with the tracks belonging to it. Among them, we use the vertex mass (see Sec.~\ref{s:sc}), 
defined as the invariant mass of all tracks originating from the secondary vertex;
in the computation of the vertex mass, tracks are assumed to be charged pions 
($M_\pi=139$ MeV/$c^{2}$), there being no possible discrimination available between different
particle species.

A jet is called {\em taggable} if it contains at least two tracks 
that pass all the selection criteria applied by the SecVtx algorithm 
other than the large impact parameter requirement. A jet is
said to be {\em $b$-tagged} if a secondary vertex with good fit
quality is found by the SecVtx algorithm and if the angle between the jet direction and the vector pointing from 
primary vertex to the secondary vertex is less than $\pi/2$.
So-called {\em negative tags} are those failing the latter requirement: these
are mostly light-quark or gluon jets with a fake secondary vertex
due to imperfect track resolution. Tagged jets constitute a subclass of taggable jets.

SecVtx has been demonstrated to efficiently tag $b$-quark jets (40-50\% efficiency for $b$-jets from top quark decay), with
light-quark and gluon fake rates of less than 1\%~\cite{cdfdet}.
For the extraction of a $Z$ boson decay signal, whose cross section
is smaller by more than three orders of magnitude than that of
generic dijet background, it is mandatory to maximize the $b$-jet
content of the data sample. Consequently both jets are required to
be tagged by SecVtx. In the next section we show that this
requirement is still needed for data collected with an impact parameter
trigger.

\subsection{Heavy flavor content of the data}\label{s:sc}

The trigger described in Sec.~\ref{s:trigger}
produces a dataset containing a significant fraction of events with
two central $b$-quark jets. However, it is only through the direct
reconstruction of the secondary vertex inside each of the two
central jets that we can remove most of the light-quark and gluon
background, because its production cross section is so large that it
still dominates the data after trigger selection.

\begin{figure}[h!]
\centerline{\includegraphics[width=12cm]{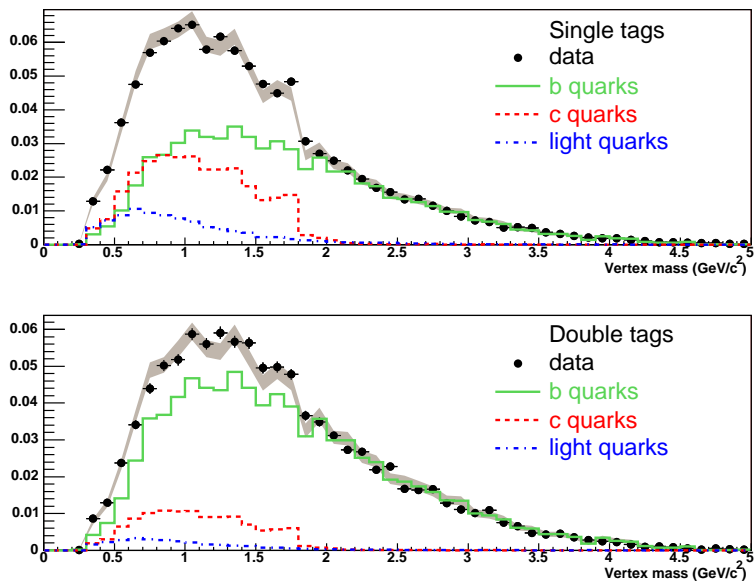}}
\caption {\em Invariant mass of the charged tracks in the vertex for
a sample of jets in
              single-tagged events (top) and for the tagged jet in double-tagged
              events (bottom). The templates show the relative fraction of $b$,
              $c$, and light quark or gluon jets estimated from the sample composition fit.
\label{f:btagdistr}}
\end{figure}

We determine the fraction of events due to $b\bar{b}$ production
using the vertex mass, which is sensitive to
the flavor of the parton originating the jet: light quarks and
gluons, which generate a secondary vertex tag only by virtue of
track mismeasurement, produce vertices with low invariant mass on average;
charm and bottom quarks have vertices with larger mass, and the
latter is easily distinguishable from the former (see
Fig.\ref{f:btagdistr}).

For this study, the selection of clean dijet events was made tighter
with respect to the initial data to ensure a better understanding of
the event topology. Events are selected if they contain two jets in
the central calorimeter ($|\eta|\leq1.0$) with uncorrected $E_T>20$
GeV and if there are no jets with uncorrected $E_T>10$ GeV in the $|\eta|>1.0$ calorimeter regions.
The data are divided into subsets based on the trigger settings and
running conditions in order to check for variations in the sample
composition. 

The results show that the mean fraction of events due to direct
production of a central $b\bar{b}$ pair in our data sample is
$F_{bb}=23 \pm 2\%$ in clean dijet events before SecVtx tagging
requirements; $F_{bb}=46\pm5\%$ in events with one jet containing a
SecVtx tag; and $F_{bb}=91^{+9}_{-10}\%$ in events with both jets SecVtx
tagged.
These studies therefore indicate that in order to collect a sample
with high $b\bar b$ purity it is necessary to require secondary
vertex tagging of both central jets. 

\section{Signal selection}\label{s:optim}

In this section we define the dijet system used for the
reconstruction of the $Z$ mass peak and the kinematic selection cuts
we apply to increase the signal-to-background ratio.

\subsection {Preliminary cuts and definition of the dijet system} \label{s:dijet}

As we discussed in Sec.~\ref{s:data}, our initial dataset consists
of events that pass a level 2 trigger requirement of two central
calorimeter clusters with a very low $E_T$ threshold, which was set at 3 GeV
to affect as little as possible the shape of the turn-on in the
dijet mass distribution. The low threshold is needed because the
$E_T$ of soft jets is measured rather poorly by the level 2 trigger
hardware cluster finder~\cite{pacman}.
In the level 3 trigger, jets are required to have $E_T>10$ GeV
using the standard CDF jet cone algorithm.  The dijet mass turn-on is
significantly affected by this requirement because the jet energies
are still uncorrected. However this requirement is mandatory to
keep the trigger rate at an acceptable level. We must carefully
model the effect of the trigger selection on the mass distribution.
This is made easier if the region of maximum variation in the jet
reconstruction efficiency, the turn-on region, is discarded by means
of a sharp offline cut.

We studied the raw and corrected $E_T$ distributions of
the two leading jets in our data in order to select an initial
threshold to define our dijet sample. The jets we considered have
detector pseudorapidity in the range $|\eta_{d}| \le 1.0$, a
selection that reflects the cuts applied by the trigger
and which limits threshold effects in the dijet mass distribution.

\begin{figure}[t!]
\begin{minipage}{0.49\linewidth}
\centerline{\includegraphics[width=7.5cm]{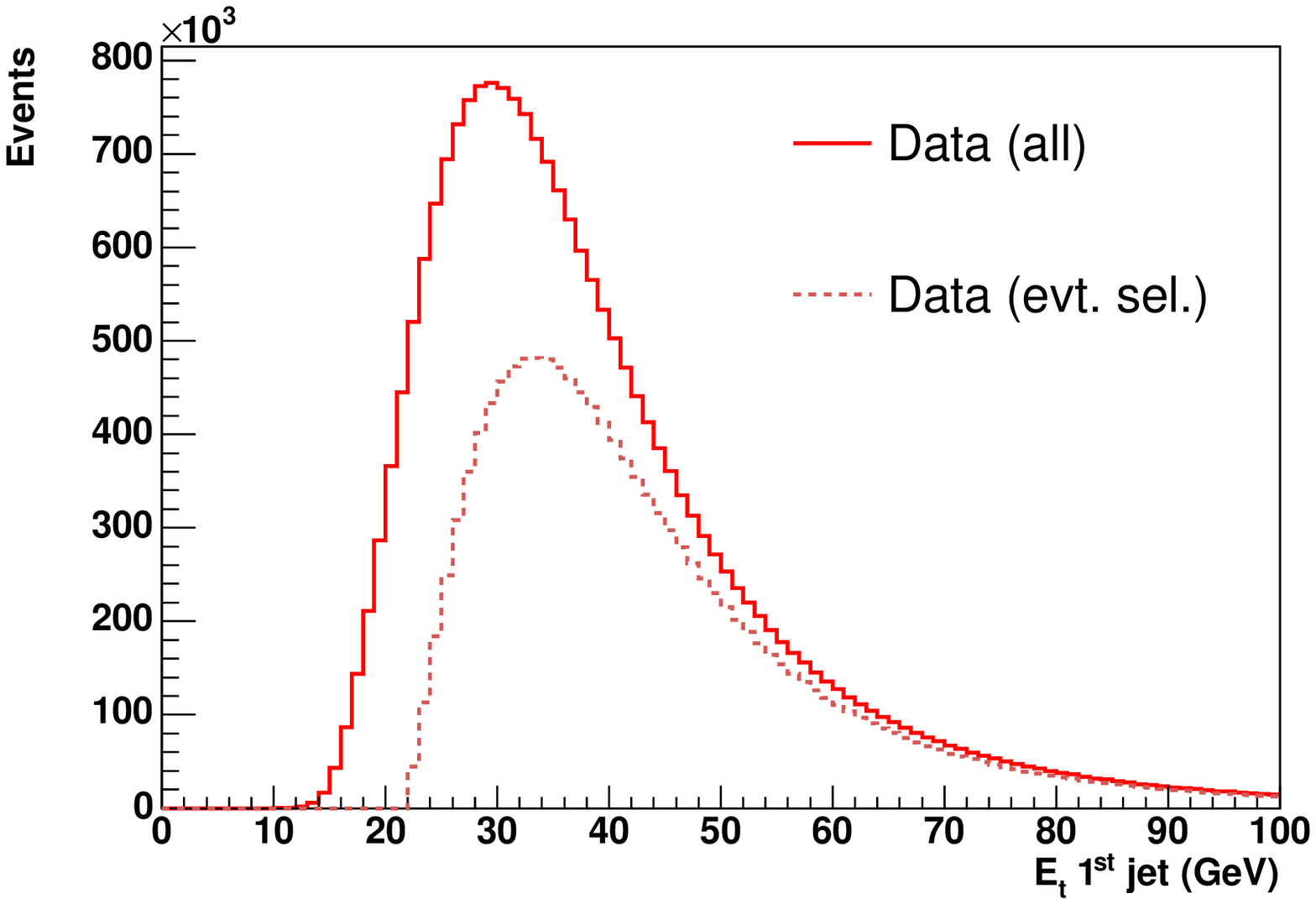}}
\end{minipage}
\begin{minipage}{0.49\linewidth}
\centerline{\includegraphics[width=7.5cm]{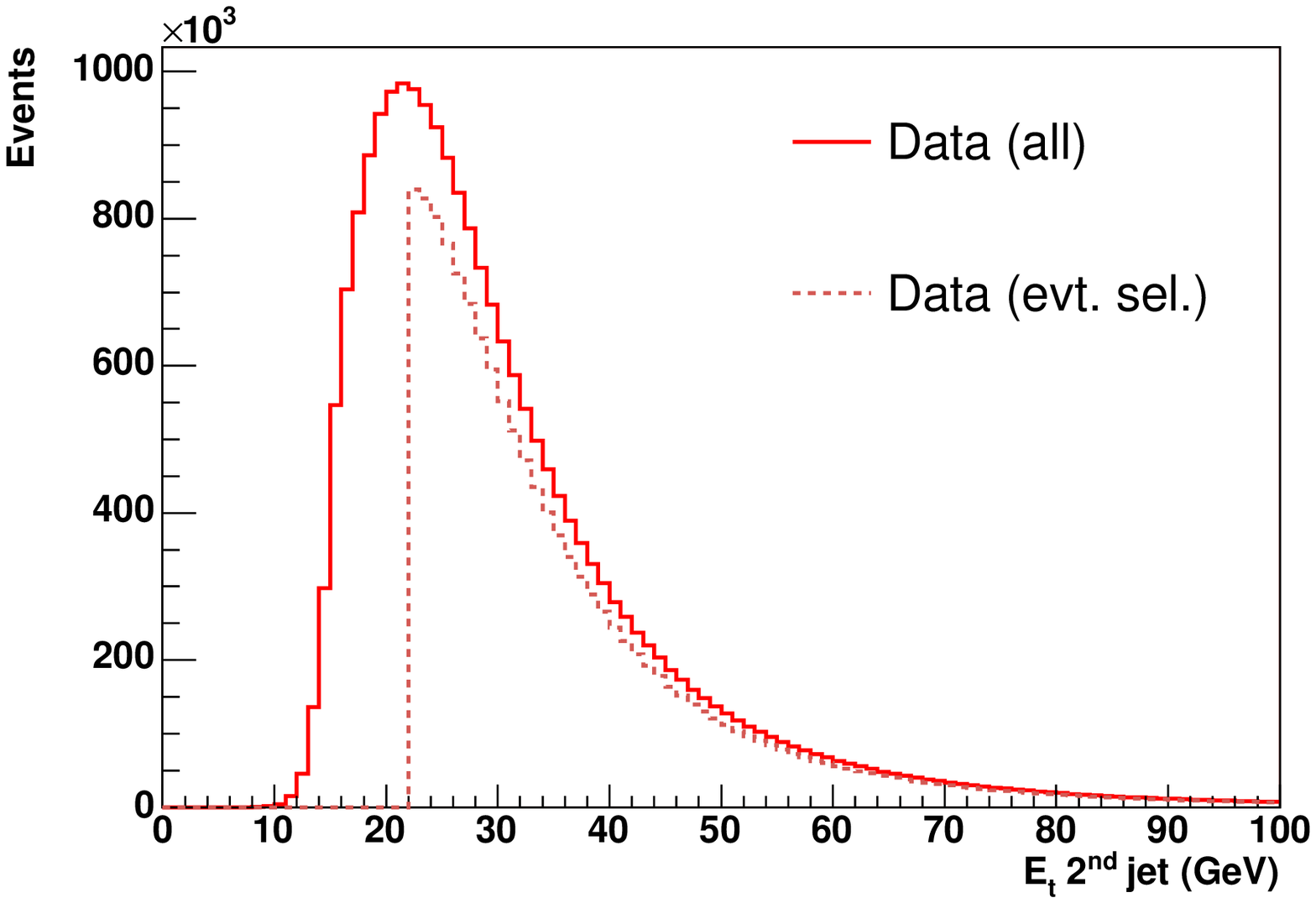}}
\end{minipage}
\caption {\em Jet $E_T$ distributions for the two leading jets in experimental
           data. Left: corrected $E_T$ of the leading jet for all events (continuous line)
           and events passing the preliminary selection (dashed line). Right: same, for the second
           jet. \label{f:etturnons_ets}}
\end{figure}

Given our goal of determining a scale factor for the
energy of jets corrected to level 5 by means of a fit to the $Z\to b\bar{b}$ signal, 
the most sensible way to define the dijet system is to use jet energies corrected to that level. 
Based on Fig.~\ref{f:etturnons_ets}, we require that both jets have corrected
$E_T>22$ GeV. We thus select events unbiased by the rise of
the trigger turn-on curve.
Below is the step-by-step procedure adopted to define the central dijet
system used to compute the dijet mass.

\begin{enumerate}
\item Reconstruct jets with the JetClu algorithm using a cone radius $R=0.7$;
      correct the energy of jets to level 5;
\item order the list of jets by decreasing value of $E_T$; require that the first two
      jets have $E_T>22$ GeV;
\item select central dijet events by requiring that the two leading jets
      have detector pseudorapidity in the range $|\eta_d| \leq 1.0$;
\item require both leading jets to be defined as taggable by SecVtx.
\end{enumerate}

\noindent Table~\ref{t:stats_prelim} details the number of events
passing these cuts.

\subsection {Optimization of the kinematical selection}\label{s:optkin}

\begin{figure}[h!]
\begin{minipage}{0.49\linewidth}
\centerline{\includegraphics[width=7.5cm]{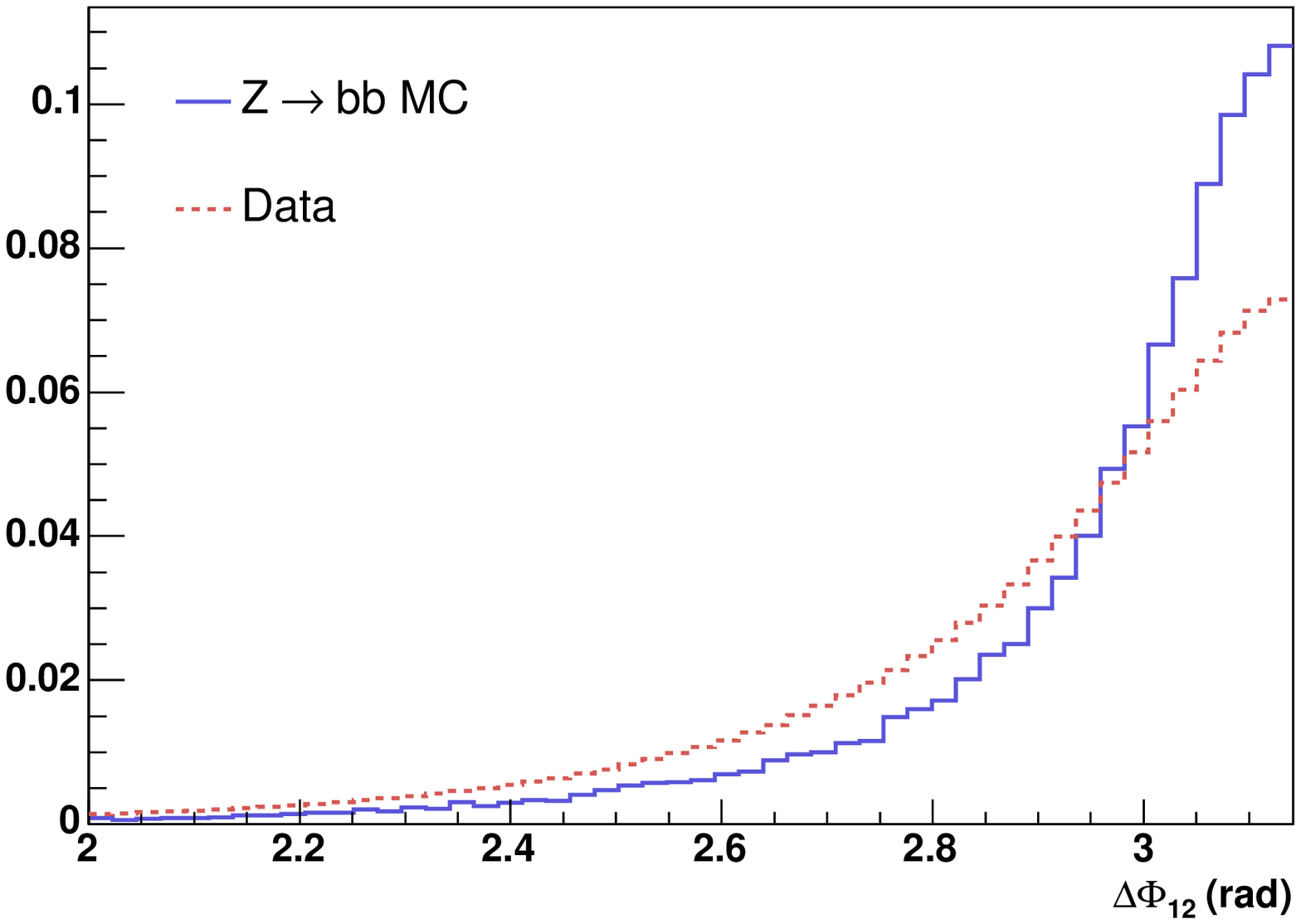}}
\end{minipage}
\begin{minipage}{0.49\linewidth}
\centerline{\includegraphics[width=7.5cm]{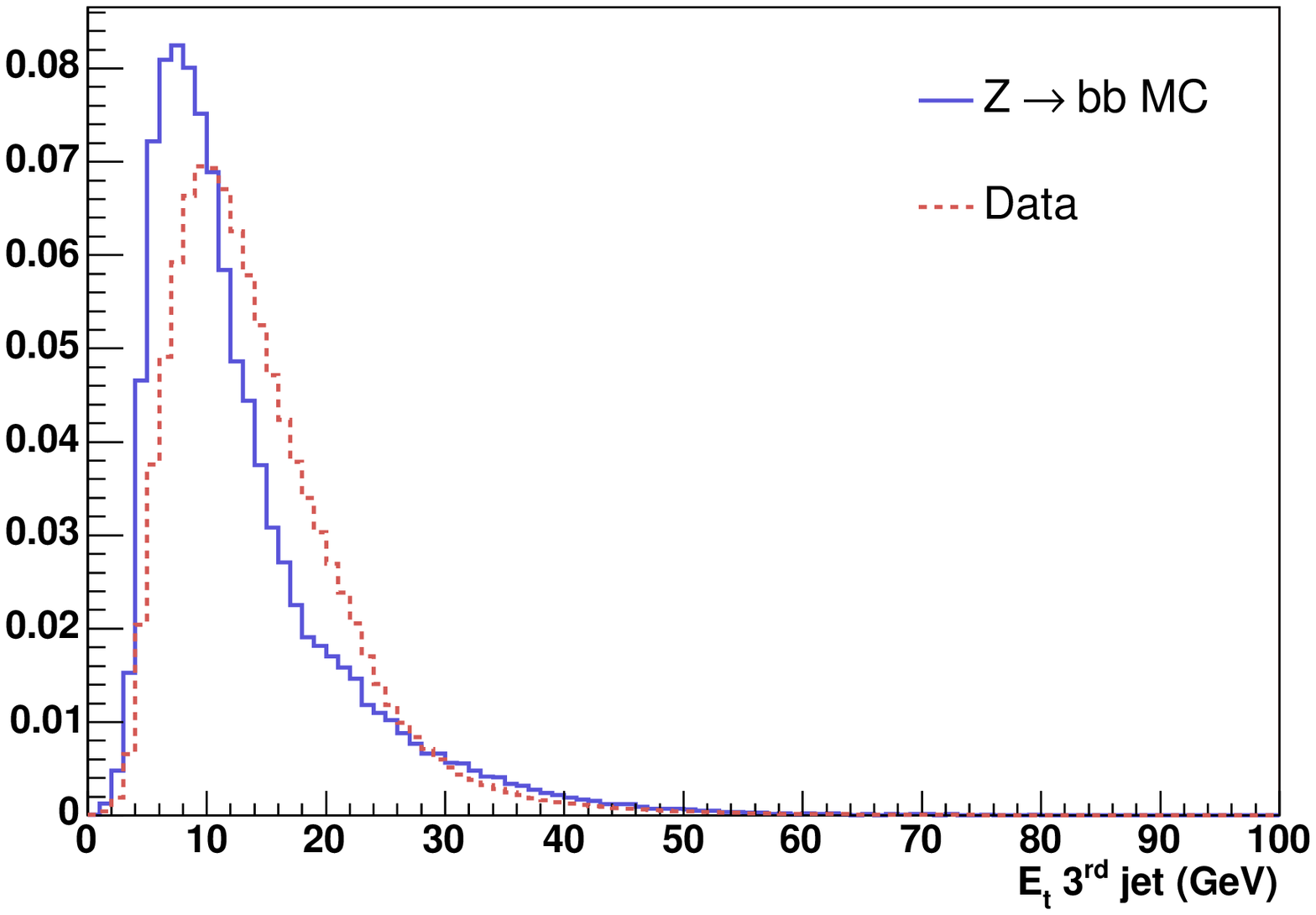}}
\end{minipage}
\caption {\em Distributions (normalized to unity) of the variables used to optimize the kinematical selection, for data and Monte Carlo (Pythia).
          Left: azimuthal angle between the leading jets; right: corrected $E_T$ of the third jet. \label{f:kinvars}}          
\end{figure}

The selection of $Z$ bosons that decay to
$b$-quark jets and their discrimination from the huge QCD background
is a difficult experimental problem. Background events  differ only
slightly from the signal: they feature a higher probability of gluon radiation
from the initial state, a different color configuration of the final state,
and of course a non-resonant structure in the invariant mass. 
We studied many kinematic variables which could in principle be sensitive to these
differences. However, very little can be done after selecting a
clean dijet system in which both jets are $b$-tagged. The most sensitive
quantities providing additional discrimination are the transverse
energy of the third jet, $E_T^3$, and the angle between the leading
two jets in the plane transverse to the beam, $\Delta \Phi_{12}$. A
comparison of data and Monte Carlo for these variables is shown in
Fig.~\ref{f:kinvars}.

To determine the best cuts on these two variables we employ
pseudo-experiments, which allow us to estimate the $b$-jet energy
scale uncertainty we would obtain from a fit to the dijet mass
distribution with the number of events and signal fraction expected
for a given set of kinematic cuts ($\Delta \Phi_{12}>x, E_T^3<y$, where $x$ varies from 2.0 to 3.1 radians and $y$ varies from 1 to 25 GeV).

We construct pseudo-data distributions of the dijet invariant mass from a data-driven background template (see Sec.~\ref{s:bgr_sig_templ}) and a $Z \to b\bar{b}$ MC signal template corresponding to a $b$-jet energy scale $k=1.0$. For each pseudo-experiment, we perform a simple two-components $\chi^2$ fit. We use as a
background template in the fit the same distribution used to
generate the background mass values in the pseudo-data distribution, while for the signal we  use in turn 21
different templates generated at varying values of the $b$-jet
energy scale $k$ from 0.9 to 1.1. The width of the resulting
$\chi^2$ curve gives us the accuracy of the fit.

To first order, the minimum uncertainty in the $b$-jet energy scale
among a set of fits with varying data statistics and signal to noise ($S/N$) 
values should correspond to the largest value of $S/\sqrt{N}$. However,
we must also consider a systematic uncertainty on the
background shape. A priori studies of the accuracy of our background
modeling performed on background-enriched data events where only one jet is $b$-tagged (very low signal fraction: $S/N\sim0.2\%$)
allow us to predict a shape uncertainty of $1\%$. 
For these pseudo-experiments, therefore, we include the effect of a background uncertainty in the fits by adding in
quadrature an uncertainty equal to $1\%$ of the bin content to the
uncertainty of each bin in the pseudo-data distribution. The effect of an inflation by
$1\%$ in the statistical error is not exactly the same as a
systematic uncertainty in the knowledge of the background shape, but
it is a good first-order guess of the impact that a similar
uncertainty has on a fit to the dijet mass distribution with a given
signal fraction.  This is used only to make our a priori
choice of the $E_T^3$ and $\Delta \Phi_{12}$ cut values.
Pseudo-experiments show that the minimum uncertainty on the
$b$-JES can be obtained by selecting the data with cuts at $E_T^3 <
15$ GeV and $\Delta \Phi_{12} > 3.0$ radians. Those are the cuts we
use to define the region in which we search for $Z \to b\bar{b}$
signal. 

The total data with two central $b$-tagged jets passing the preliminary selections and the 
$\Delta \Phi_{12}$ and $E_T^3$ cuts amounts to $267\ 246$ events. The predicted signal fraction
in this sample is about $1.7\%$. A more precise estimate of this fraction is provided in the next section.

\section{Signal acceptance and related uncertainties}\label{s:effsig}

In this section we estimate the amount of $Z\to b\bar{b}$ signal in
the selected data. That number is an input to fits we
describe in Sec.~\ref{s:likfits} to extract the $b$-jet energy scale factor
from the data, because a signal normalization constraint helps
reduce the fit uncertainties. We also evaluate in this section the
contributions from different sources of contamination to our sample.
Finally we discuss the impact of our modeling of initial state
radiation on signal acceptance uncertainty.

\subsection{Number of events in the signal region}
\label{s:nsig}

Monte Carlo events of $Z \to b\bar{b}$ signal were required to pass
the Z\_BB trigger simulation and the same offline cuts applied to
the experimental data.
Multiplying the signal efficiency obtained from MC to the cross section for $Z$ boson production times the 
branching ratio of $Z$ decay to $b$-quark pairs~\cite{nlozxs} and to the total integrated luminosity of the data,
we obtain a raw estimate of 9478 events with two SecVtx taggable jets passing
the kinematic selection, 4164 of them with both jets $b$-tagged.

The modeling of $b$-quark tagging in the Monte Carlo simulations
depends on subtle characteristics: physics effects such as the
admixture of $B$ mesons and baryons produced in the fragmentation,
their lifetimes, and their momentum spectra; and detector modeling
issues such as the detailed description of track position
measurements in the silicon microstrip detector and track-finding
efficiency inside jets. For these reasons, CDF measures a $b$-tagging scale factor, defined as
the ratio of SecVtx vertex-finding efficiency on $b$-quark jets in
the data to that in Monte Carlo simulated $b$-jets. The SecVtx $b$-tagging efficiency scale factor has been determined
for the run range used in the present analysis~\cite{btagSF} as:
\begin{equation}
SF(\mbox{b-tagging}) =  0.95 \pm 0.01(\mbox{stat}) \pm 0.04(\mbox{syst}).
\label{eq:btagSF}
\end{equation}
In events where both leading jets are $b$-tagged, this scale factor is squared.

For a proper estimation of the signal efficiency we determined the accuracy of 
the Z\_BB trigger requirements in data and Monte Carlo.
With that aim, the selections used in the Z\_BB trigger
can be divided in two parts: requirements on the online-measured
energy deposits in the calorimeters, and requirements on
online-reconstructed tracks. The data-MC difference in the
calorimeter-based efficiencies can be studied by comparing inclusive
jet samples of data and simulation, while the differences in track
requirement efficiencies can be studied in a subset of jet samples
enriched in $b$-quark jets. The data-MC difference in the trigger
efficiency can then be expressed as

\begin{center}
$SF[\epsilon_{data}/\epsilon_{MC}]
=SF(\mbox{calor-trigger}) \times SF(\mbox{track-trigger})$.
\end{center}

By comparing jet samples in data and Monte Carlo we found a data/MC ratio of
$SF(\mbox{calor-trigger})=1.10\pm0.01(\mbox{stat})\pm0.01(\mbox{syst})$ in events containing
two jets with energy similar to those of $Z$ decay. For tracking requirements, the data/MC ratio was measured to be
$SF(\mbox{track-trigger})=1.12\pm0.06(\mbox{stat})\pm0.08(\mbox{syst})$ by using
dijet events where both jets contained a signal of $b$-quark decay.

Using the numbers described above, we can now compute our estimate
for the number of events with two positive SecVtx $b$-tags expected
in the signal region defined in section \ref{s:optkin}:
\begin{eqnarray*}
N_{exp}^{++} &=& N_{exp}^{++}(\mbox{raw}) \times \left[SF(\mbox{b-tagging})\right]^{2}
\times SF(\mbox{calor-trigger}) \times SF(\mbox{track-trigger}) \\
             &=& 4164 \times (0.95)^2 \times (1.10) \times (1.12) = 4630 \ \textrm{events.}
\end{eqnarray*}

\noindent The various sources of uncertainty affecting this estimate are
discussed in the remainder of this section.

\subsection{Initial state radiation uncertainty}
\label{s:SysUncISR}

The kinematic selection applied to our data makes the expected
number of signal events dependent on the jet activity in the event,
and in particular on the modeling of initial state QCD radiation
(ISR).

To estimate the acceptance systematics due to the modeling of ISR, we
studied a sample of $Z \rightarrow e^{+} e^{-}$ and
$Z \rightarrow \mu^{+} \mu^{-}$ events collected by
high-$P_T$ electron and muon triggers in the same run range. We thus
in effect substitute electrons (muons) for the $b$-tagged jets, but
we include cuts that are used in the $Z \rightarrow
b\overline{b}$ analysis in order to mimic the selection biases our
data is subjected to: for instance, the jet $E_{T}^{3}$ cut is
applied to the leading jet of the $Z \rightarrow e^{+}
e^{-}$ ($\mu^{+} \mu^{-}$) events.

\begin{figure}[htb]
\begin{minipage}{0.49\linewidth}
\begin{center}
        \includegraphics[width=7.1cm]{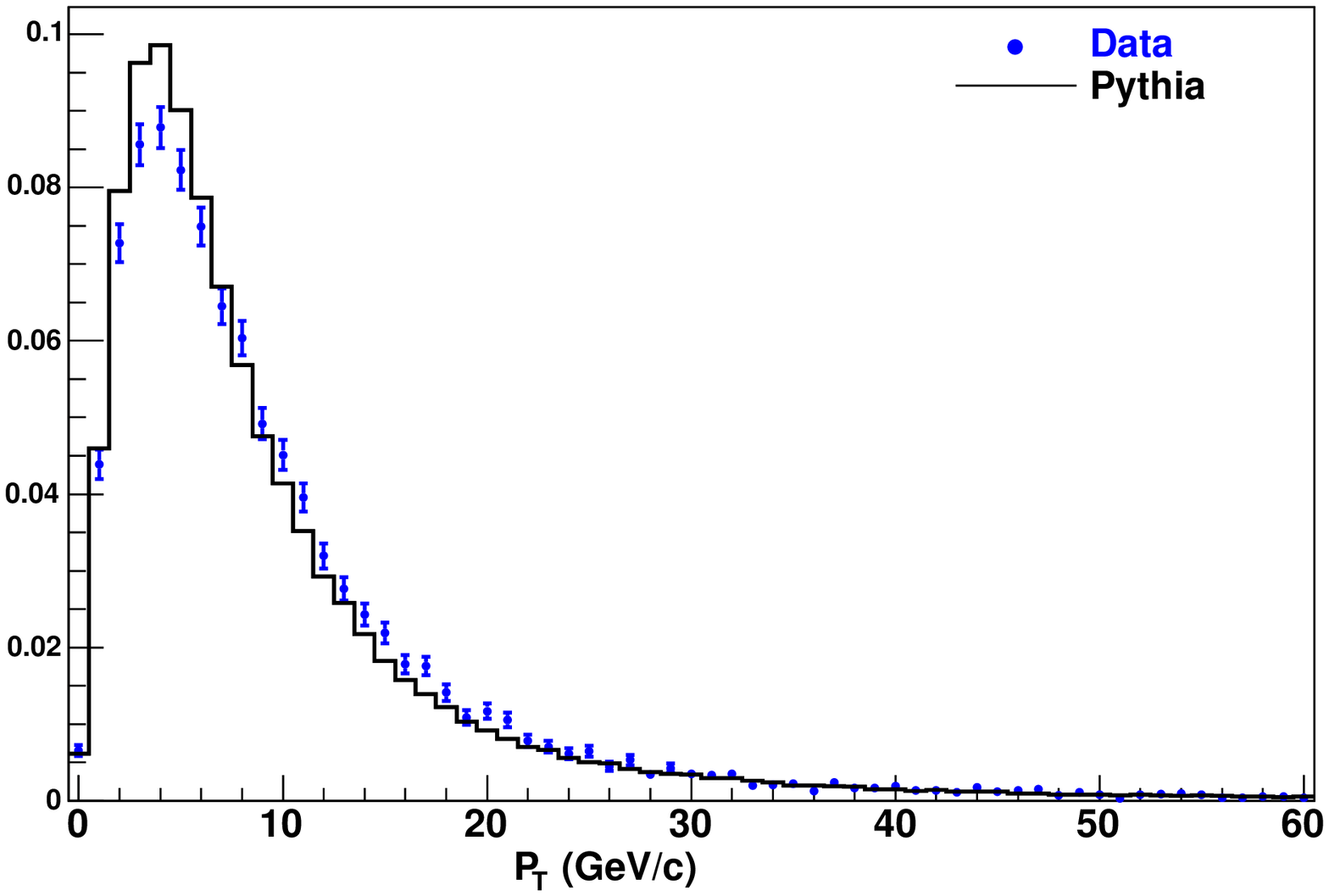}
\end{center}
\end{minipage}
\begin{minipage}{0.49\linewidth}
\begin{center}
        \includegraphics[width=7.1cm]{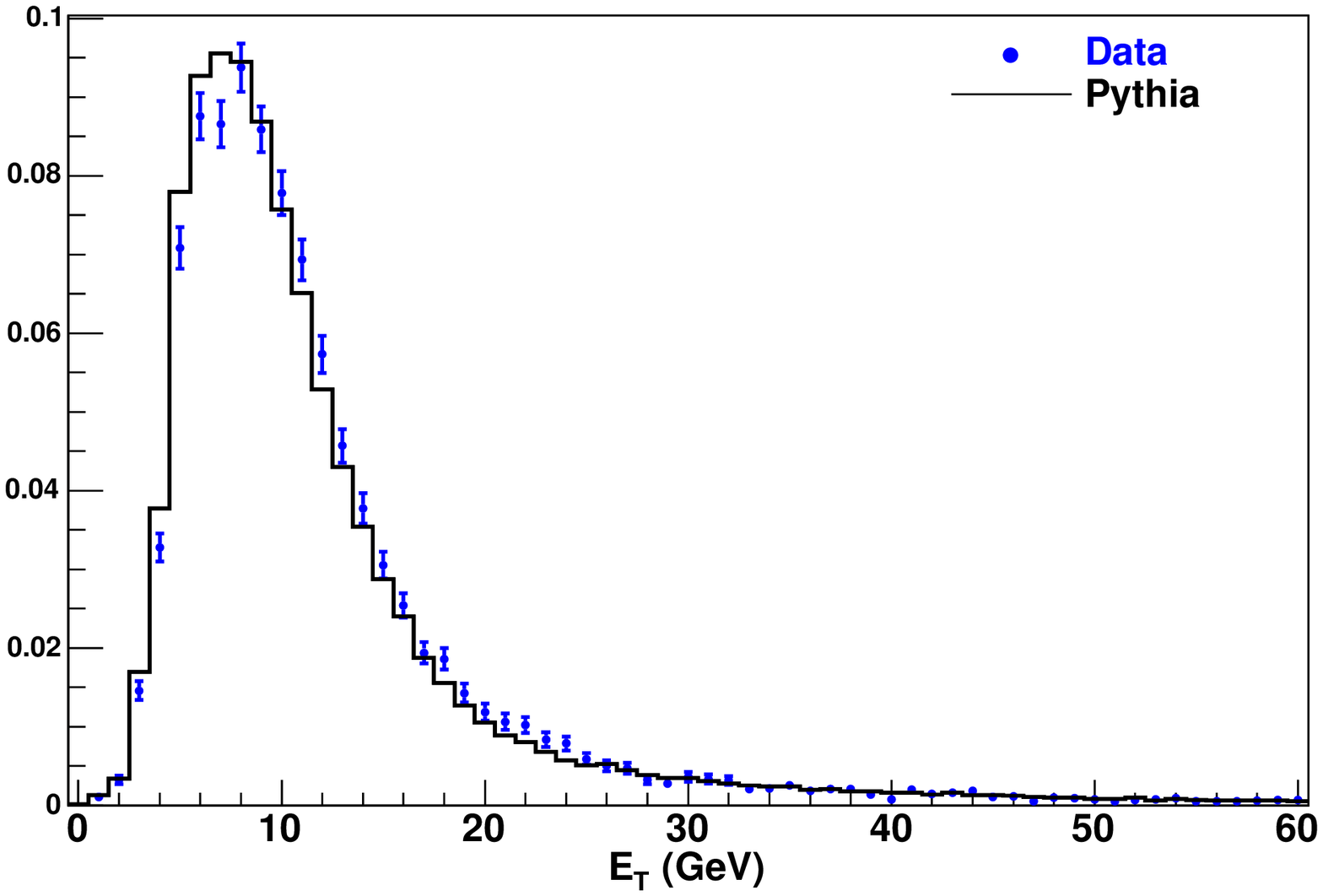}
\end{center}
\end{minipage}
\caption{{\em Left: $P_{T}$ distribution of the $Z$ boson for electron
data and MC. Right: corrected $E_{T}$ distribution of the leading jet for electron data and MC. All distributions are normalized to unity.}}         
\label{f:cZpt_060}
\end{figure}

\begin{figure}[htb]
\begin{minipage}{0.49\linewidth}
\begin{center}
        \includegraphics[width=7.1cm]{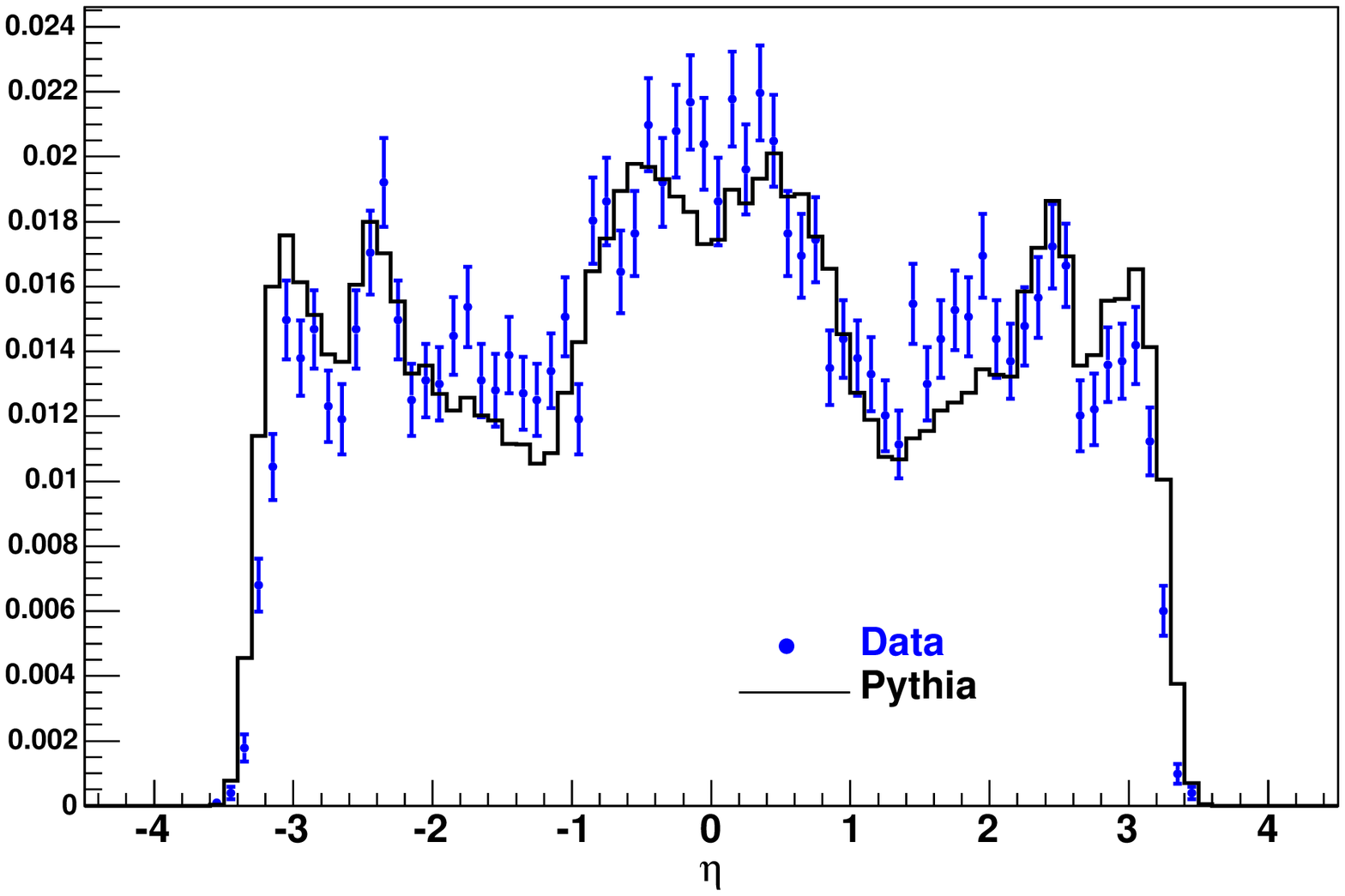}
\end{center}
\end{minipage}
\begin{minipage}{0.49\linewidth}
\begin{center}
        \includegraphics[width=7.1cm]{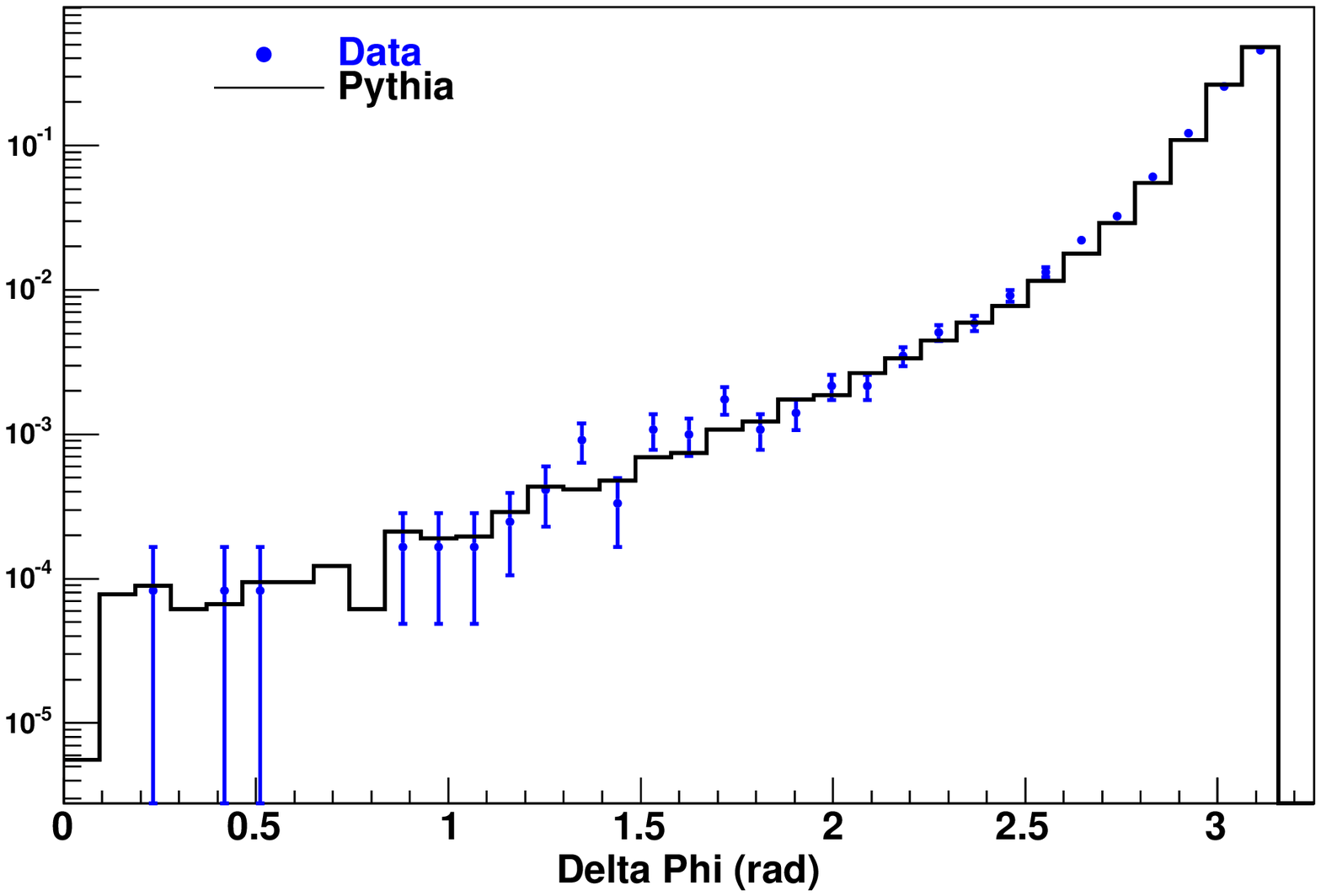}
\end{center}
\end{minipage}
\caption{{\em Left: $\eta$ distribution of the leading jet for electron data and MC.
         Right: $\Delta \phi$ distributions of the electron pair. All distributions are normalized to unity.}}
\label{f:cjetEta_cut}
\end{figure}

Large samples of Monte Carlo events were produced using Pythia version 6.216
to compare to leptonic $Z$ decays in the data. The
application of a standard selection~\cite{Zsel} produces a total of 14 507 $Z
\to e^{+}e^{-}$ candidates and 13 727 $Z \to \mu^{+} \mu^{-}$ candidates in
experimental data, while the Monte Carlo samples contain $249\ 540$
and $427\ 495$ events, respectively.
Figs.~\ref{f:cZpt_060}~and~\ref{f:cjetEta_cut} show the $P_T$ of the
$Z$, the leading jet corrected $E_{T}$, the leading jet $\eta$
and the $\Delta\phi$ of the electron pair after the $Z$ cuts, for both
the electron data and MC.

We apply to the leptonic $Z$ samples kinematic cuts corresponding
to those we use to select our signal sample of $Z \rightarrow b\overline{b}$
data: the level 5 corrected $E_{T}$ of the
leading jet (mimicking the third jet in the $Z \to b\bar b$ dataset)
is required to be less than 15 GeV, and the $\Delta \Phi$ between
the two leptons is required to be greater than 3.0 radians.

In order to calculate the ISR systematic uncertainty on signal
acceptance, we re-weight the MC events so that they perfectly match
the distribution of electron and muon data in three different
distributions: the $P_{T}$ of the $Z$, the corrected $E_{T}$
of the leading jet, and the $\Delta \Phi$ between the two leptons.
By dividing the difference between the weighted and unweighted MC
events by the unweighted MC events, we determine the systematic
uncertainty on the acceptance. Because these measurements are highly
correlated with each other, we take the largest of these
values, 4.6\% for the electrons and 3.0\% for the muons, and
average them together.  This average, 3.8\%, is our final estimate
for a systematic uncertainty on the acceptance of our kinematic
selection resulting from ISR.

\subsection{Uncertainty on the expected signal}
\label{s:NsigError}

There are several independent sources of uncertainty on our estimate of the number of
$Z\to b\bar{b}$ events:

\begin{itemize}
\item the statistical uncertainty on the signal efficiency calculated from MC (due to the finite sample): $1.0\%$;
\item the uncertainty on the integrated luminosity of our dataset: $5.9\%$;
\item the uncertainty on the data/MC $b$-tagging scale factor is evaluated from Eq.(\ref{eq:btagSF}):
      it amounts to $4.3\%$, and it effectively doubles since we require two $b$-tags per event
      in our analysis;
\item the uncertainty in the acceptance of the Z\_BB trigger is described in Sec.~\ref{s:nsig}.
      The effect is divided into calorimeter-cut modeling uncertainties and track-cut 
      modeling uncertainties, which respectively amount to $1.8\%$ and $8.9\%$;
\item the acceptance uncertainty from ISR modeling in the Monte Carlo was estimated in the previous
      section. This amounts to an additional $3.8\%$ systematic uncertainty.
\item uncertainty related to the modeling of final state radiation (FSR): by varying the
      parameters governing QCD radiation of final state $b$-quarks in $Z$ Monte Carlo events
      an additional $2.9\%$ uncertainty is estimated.
\end{itemize}

The above sources of uncertainty on the signal acceptance are summarized in table
\ref{t:SigAccUncertainty}. The total quadratic sum of these uncertainties amounts
to 14.7\%. We thus expect $N_{exp}^{++} = 4630  \pm 681$ events
of signal to pass all the steps of our event selection.


\subsection{Residual contaminations from other physics processes}\label{s:rescont}

We have defined, in section \ref{s:optkin}, the signal region as the sample of events with two central leading $b$-tagged jets that pass our $\Delta\Phi$ and $E_T^3$ kinematical selections. On the other hand, events with two $b$-tags failing either one of these $\Delta\Phi$ and $E_T^3$ cuts, as well as events with two taggable jets (that pass or fail the kinematical selections) are background-enriched regions that are used to construct the background shape used in the fitting procedure (see section \ref{s:templates}).

All these samples are dominated by the large background from generic QCD $2 \to 2$
scattering reactions. However a few additional physics processes can contaminate
this data. Among these are $Z \to b\bar{b}$ events leaking into the
background regions, and $Z\to c\bar{c}$ and $W\to c\bar{s}$ events
that pass our selections since $c$-quarks can produce a secondary
vertex.

$Z$ decays present in the signal zone defined by the cuts $\Delta\Phi>3.0$ and $E_T^3<15$ GeV
amount to about $1.7\%$ in events with two $b$-tagged jets, while the fraction is $0.2\%$ in
events with two taggable jets. Among events failing the kinematical cuts, 
the fraction of $Z \to b \bar b$ decays is
equal to $0.8\%$ in events with two $b$-tagged jets, and falls by an order of magnitude
($0.08\%$) in events with two taggable jets. The signal contamination in samples 
used to construct our background model will
be accounted for in section \ref{s:templates}, since a fraction of signal
higher than 0.10\% starts to affect appreciably the shape of the dijet mass
distribution and thus the modeling of our background function.

Samples of $Z\to c\bar{c}$ and $W\to c\bar{s}$ MC events have been studied
 to evaluate the contamination of these physics processes in our data. For the 
$Z\to c\bar{c}$ channel we expect, in data with two taggable
jets, a contamination below 0.02\% for events that pass the
$\Delta\Phi$, $E_T^3$ kinematical selections and a contamination of about
0.01\% for events that fail these selections. The contamination of
this process in data with two tagged jets is 0.03\% for the signal
region and about 0.02\% in the background region. Similarly, for $W\to c\bar{s}$ 
channel we expect, in events with two taggable jets,
a contamination below 0.05\% for events that pass the kinematical selection and
a contamination of about 0.03\% for events that fail this selection.
The contamination in events with two tagged jets is negligible (below 0.01\%). 
Each of these contributions is smaller than the smallest contribution coming 
from $Z\to b\bar{b}$ process. In summary, the $Z\to c\bar{c}$ and $W\to c\bar{s}$
decays have a negligible impact on the determination of $b$-jet energy scale factor, and
we do not correct our data for the presence of these processes.

\section{The fitting procedure}\label{s:templates}

In this section we describe the parametrization of signal and background
templates and the unbinned likelihood fit method we employ to extract the
$Z$ signal and measure the $b$-jet energy scale factor.

\subsection {Construction of background and signal templates \label{s:bgr_sig_templ}}

\subsubsection {Background templates}\label{s:BGscan}

To model the dijet mass shape of the background collected in double $b$-tagged data
after the selection of events with $\Delta \Phi_{12}>3.0$ and $E_T^3<15$ GeV
we rely on a three-step procedure:

\begin{enumerate}
\item
we first determine the invariant mass shape of the ratio $R(m_{jj})$
between events with two positive SecVtx $b$-tags
(labeled ``(++)'') and events with two taggable jets
(labeled ``(00)''), in several regions with poor signal fraction. Such regions are defined by 
selecting events that fail one or both of the cuts $\Delta \Phi_{12}>x$ and $E_t^3<y$ (with $x \leq 3.0$, $y\geq 15$ GeV). 
These background regions have by construction zero overlap with the region
of kinematical space where we look for the signal. $R$ is thus defined as

\begin{eqnarray}
R(m_{jj};x,y) = \frac{N_{++}(m_{jj};x,y)}{N_{00}(m_{jj};x,y)}; \nonumber
\end{eqnarray}

\item the ratio $R$ is then multiplied by the mass distribution of events with two 
taggable jets found within the signal region ($\Delta \Phi_{12}>3.0$, $E_T^3<15$ GeV) 
to construct a background template as a function of the two parameters $x,y$;

\item finally, the resulting distribution is fit with a continuous parameterization 
that well models the full spectrum of invariant masses between 0 and 200 GeV/$c^{2}$.
\end{enumerate}

\begin{figure}[h!]
\centerline{\includegraphics[width=10cm]{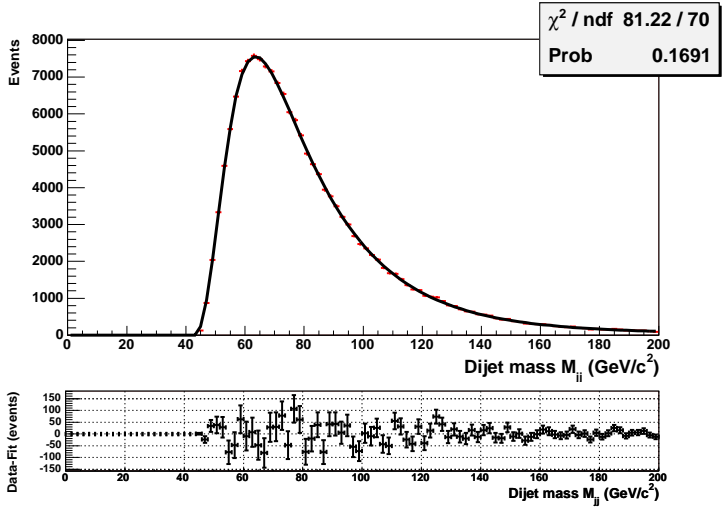}}
\caption {\em Example of a data-driven background dijet mass distribution fit
with a Pearson IV function plus an error function.
          \label{f:BGPDFex}}
\end{figure}

In order to describe the data-driven background shape, derived as discussed above,
we choose as a probability density function (p.d.f)
the sum of a Pearson IV function~\cite{Pearson} and an error function (erf):
\begin{eqnarray}
P_b(m_{jj}) & = & \beta_0 \left[ 1 + \left(\frac{m_{jj}-\beta_1}
{\beta_2}\right)^2 \right]^{-\beta_3} \cdot e^{-\beta_4 \ \tan^{-1}\left
(\frac{m_{jj}-\beta_1}{\beta_2}\right)} \nonumber \\
& & + \beta_5 \times \textrm{erf} \left( \frac{m_{jj}-\beta_6}{\beta_7} \right) + \beta_8,
\label{eq:bgshape}
\end{eqnarray}
where $\beta_i$ are the parameters of the p.d.f (see an example in Fig.~\ref{f:BGPDFex}).

If we select a background region too close to the signal region, it
will be appreciably contaminated with signal events, while a
background region based on extreme values of the parameters (very
low $x$ or high $y$ for example) will
select events kinematically quite different from those populating the signal
region and will likely fail to provide a satisfactory model of the background in
the signal region. Understanding the correlation between the selection
of the background region and the tag rate function $R$ is not trivial
since the tag rate shape depends not only on the kinematic
variables but also on the sample composition of data in the
background region. We thus have no meaningful way to favor a
priori a background model, that is, a particular choice of 
($x,y$) over another. We thus consider
a large set of background models in our fitting procedure.

We construct different background models by scanning the $(x,y)$ variables space: 
we vary the value of $x$ from 2.50 to 3.00 rad in steps of 0.02 rad and we vary 
$y$ from 15 to 25 GeV in steps of 1 GeV. This yields a total of 286 possible forms of the ratio $R$. All
these determinations are correlated to each other, but they possess
slight differences. These differences translate into 286 different
background templates.

The contamination of $Z\to b\bar{b}$ signal in the 286 background regions
must be accounted for (see section \ref{s:effsig}). There is a 0.6-0.8\%
signal fraction in events with two $b$-tagged jets,
depending on the $(x,y)$ selection used.
The fraction of signal in taggable dijets in the background regions is instead
a factor of ten smaller. We correct the signal contamination by estimating from Monte Carlo
the amount of $Z\to b\bar{b}$ signal in the background region for
each choice of kinematic cuts, and by
subtracting the expected signal from each dijet invariant mass
template used to construct the tag rate function $R$.  The mass
distribution of events with two taggable jets in the signal region,
to which the tag rate function is multiplied to construct the
background shape, also contains a small contamination of signal, and
is corrected accordingly.
A data/MC $b$-JES factor of 1.00 is
assumed when subtracting MC templates from the data. A systematic
uncertainty on the final measurement will be estimated later to account
for this specific choice (see section \ref{s:systs}).

\subsubsection {Signal templates}

\begin{figure}[h!]
\centerline{\includegraphics[width=10cm]{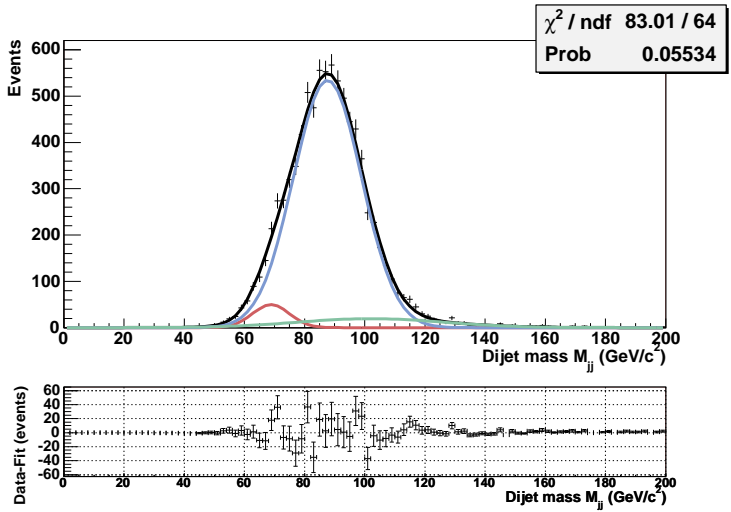}}
\caption {\em Signal dijet mass distribution fitted with a sum of
three gaussian functions (for k=1.0).
          \label{f:exsignalPDF}}
\end{figure}

To construct dijet mass signal templates we use distributions of fully
simulated Monte Carlo $Z \to b\bar{b}$ events. In events
that pass the Z\_BB trigger simulation we apply a factor $k$ to the 
energy of each jet in order to mimic a data/MC scale factor; we then apply
to modified jet energies the same event and kinematic selection applied 
on the data. As $k$ varies from 0.90 to 1.10 in steps of 0.01 we can thus construct 21
different dijet mass templates for the signal. Each of these
distributions is fit to a sum of three gaussian functions, as
shown in Fig.~\ref{f:exsignalPDF}. To obtain one single probability
density function which has dijet invariant mass and $b$-jet
energy scale as parameters, $P_s(m_{jj},k)$, we fit simultaneously
the 21 templates allowing each parameter of the three gaussian
functions to vary linearly with the scale factor $k$. 

\subsection {The fitting function}
\label{s:likedef}

We use an unbinned likelihood procedure to measure the number of
signal and background events (respectively $n_s$ and $n_b$) and the
$b$-jet energy scale factor $k$ in our data. We find the probability
that the observed data is described as an admixture of background
events and $Z \to b\bar{b}$ events with a data/MC $b$-jet energy scale
$k$, by employing the following likelihood function:
\begin{eqnarray}
\mathcal{L}(k) = \mathcal{L}_{shape}(k) \times \mathcal{L}_{(n_s + n_b)}
\end{eqnarray}
with
\begin{eqnarray*}
\mathcal{L}_{shape}(k) = \prod_{i=1}^{N} \frac{ n_s P_s(m_{jj};k) + n_b P_b(m_{jj}) }{ n_s + n_b }
\end{eqnarray*}
and
\begin{eqnarray*}
\mathcal{L}_{(n_s + n_b)} = \frac{ e^{-(n_s+n_b)}(n_s+n_b)^{N} }{ N! }, \nonumber
\end{eqnarray*}
where the term $\mathcal{L}_{shape}(k)$ is the product, over all events, of the
probability that the $i^{th}$ event with dijet mass $m_{jj}$ is described by
background p.d.f $P_b(m_{jj})$ and signal p.d.f $P_s(m_{jj};k)$, given a $b$-jet energy
scale ($b$-JES) factor $k$. The second term $\mathcal{L}_{(n_s + n_b)}$ is introduced to constrain
the total number of signal and background events ($n_s + n_b$) to the event
count $N$ in the selected data sample.

We minimize $-\ln(\mathcal{L})$ to find the best $b$-JES factor
hypothesis. The statistical error is given by the difference between
this scale factor and the scale factor at $-\ln(\mathcal{L}) + 0.5$.

\subsection {Test of the fitting procedure with pseudo-experiments}\label{s:testproc}

Before applying our fitting method to experimental data, we use pseudo-experiments 
to test its performance. In particular, pseudo-experiments allow us to check 
whether the closeness of the signal to the peak in the
background biases the resulting $b$-JES and number of signal events.

To perform pseudo-experiments we first construct a data-driven background
distribution as discussed in Sec.~\ref{s:bgr_sig_templ}, 
which we parameterize using the function defined previously
(Eq.~\ref{eq:bgshape}). Then for a given input signal fraction and simulated
$b$-JES factor we construct pseudo-data templates drawing $n_b$
background events from the background p.d.f and $n_s$ signal events from the
signal p.d.f ($n_b$ and $n_s$ are smeared according to
Poisson statistics for each pseudo-experiment). Pseudo-data distributions are
then fit using the unbinned likelihood procedure and the output parameters
($b$-JES and number of events of signal and their statistical errors) are histogrammed.

We construct pseudo-data distributions varying the input signal
fraction from 1\% to 3\% and the input $b$-JES factor from 0.95 to
1.05. For each selection we generate 1000 pseudo-experiments, each
containing a number of events equal to that observed in the signal region for
double tagged data ($267\ 246$ events).

\begin{figure}[h!]
\centerline{\includegraphics[width=8cm]{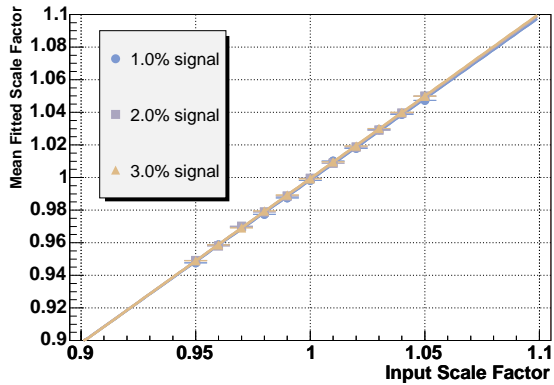}}
\caption {\em Mean fitted $b$-JES factor as a function of the input
scale factor.
          \label{f:pexp1_SFOUT}}
\end{figure}

Fig.~\ref{f:pexp1_SFOUT}
shows the mean fitted output scale factor
as a function of the input scale factor.
%
The results obtained confirm that the $b$-jet energy scale factor
can be extracted from our data even if the signal fraction is very
small and its shape peaks at a mass value not very far from the peak
of the background shape. We observe no bias from our fitting method.
However, for a check of systematic effects due to the finite
statistics of the signal template, and for an evaluation of systematic effects due to the
imprecise knowledge of the background shape, we need separate
studies. These are described in Sec.~\ref{s:systs}.

\subsection{Fits to the selected data}\label{s:likfits}

\subsubsection {Extraction of the $b$-JES and  $Z \to b\bar{b}$ signal}

The method we apply to perform the measurement of the $b$-jet energy
scale and number of events of $Z \to b\bar{b}$ signal in our
selected sample of data is the following:

\begin{enumerate}
\item we scan the $\Delta \Phi$ and $E_T^3$
kinematic space to construct 286 different data-driven dijet mass background models (see Sec.~\ref{s:BGscan});

\item each background shape is used in a preliminary binned likelihood fit of the selected data in a region of the dijet mass distribution
      containing little or no signal contamination: the ``sideband'' region
      is defined as the mass spectrum from 0 to 60 GeV/$c^{2}$ and from 120 to 200 GeV/$c^{2}$;

\item we fit the data with the unbinned likelihood function defined in Sec.~\ref{s:likedef} using in turn the different background models;

\item we choose as our measurement of the $b$-JES the result obtained with the background shape which provided
      a fit to the sideband with the highest $p$-value (step 2); Results obtained with other background shapes are used to estimate a
      systematic uncertainty related to the degree of arbitrariness of our procedure.

\end{enumerate}

\begin{figure}[h!]
\centerline{\includegraphics[width=14cm]{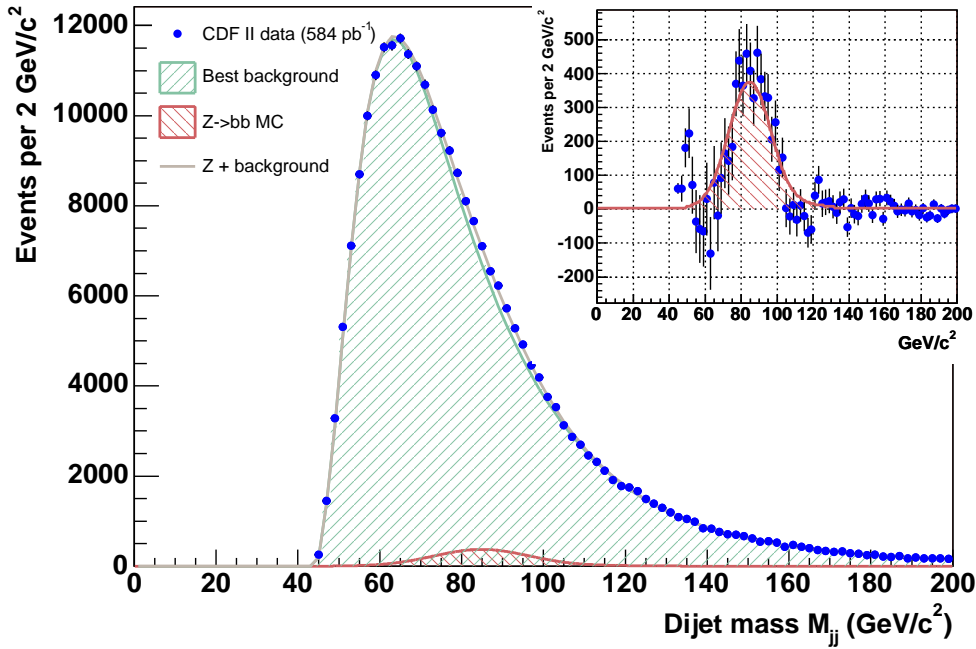}}
\caption {\em Results of the constrained unbinned likelihood fit
performed on double-tagged dijet data (points). A 
Gaussian constraint of $4630 \pm 681$ on the fitted number of signal events is
applied. The data-driven background shape and Monte Carlo signal
p.d.f are shown (hatched functions). The fit returns $5621
\pm 436$ signal events and a $b$-JES of $0.974 \pm 0.011$. The inset
on the upper right shows the data minus the background distribution
(points) and the signal shape normalized to the fitted
number of events of signal.
          \label{f:dijets_pp_FITRESULT_GC}}
\end{figure}

\noindent We increase the precision of our fit by including in the
unbinned likelihood procedure a gaussian constraint on the expected
number of signal events, computed with the Monte Carlo simulation.
We thus add to the likelihood defined in Sec.~\ref{s:likedef} a
$\exp(-\frac{(N_{sig} - n_s^{exp})^2}{2\sigma_n^2})$ term, where
$N_{sig}$ is the number of signal events and $n_s^{exp} \pm
\sigma_n$ the MC prediction for the same number.
Fig.~\ref{f:dijets_pp_FITRESULT_GC} shows the result of the
unbinned likelihood fit to double SecVtx tagged data obtained with
the background shape that best fits the sideband, when a $n_s =
4630 \pm 681$ constraint (derived in Sec.~\ref{s:effsig}) is
applied. The fit returns $N_{sig} = 5621 \pm 436$ events and a
$b$-JES factor equal to $0.974 \pm 0.011$ (errors are statistical only). The goodness of this fit
is estimated by calculating the $\chi^2/\mbox{NDF}$ corresponding to the
likelihood value at convergence, which is found to be 104/75.

\subsubsection {Additional checks}

We first check if the excess of events observed just below 50 GeV/$c^{2}$,
which is due to an imperfect modeling of the rise of the dijet mass
distribution (Fig.~\ref{f:dijets_pp_FITRESULT_GC}), 
could bias our result. We perform an unbinned
likelihood fit on the restricted mass range from 50 to 200 GeV/$c^{2}$. This
fit returns a $b$-JES factor of $0.971 \pm 0.011$ and $5862 \pm 440$
signal events. The fit has a $\chi^2/\mbox{NDF}$ of 81/70. This result is
in good agreement with the default measurement obtained by fitting
the full mass window from 0 to 200 GeV/$c^{2}$.

We then verify that the fitted $b$-JES does not vary appreciably
when the gaussian constraint on the number of signal events is
removed from the unbinned likelihood procedure. We find a fitted
$b$-JES factor of $0.971 \pm 0.011$ and a fitted signal of $6317 \pm
576$ events ($\chi^2/\mbox{NDF}=102/75$).
The signal increases by about 700 events with respect to the constrained fit,
but the $b$-JES factor remains stable within its statistical uncertainty. 

\section{Results on the $b$-jet energy scale}\label{s:systs}

In this section we discuss the sources of systematic uncertainty
affecting our determination of the $b$-jet energy scale and 
quote a complete measurement for this quantity.

\subsection {Systematics related to the background modeling \label{s:systbJES_BG}}

An evaluation of the systematic uncertainty on the $b$-JES related to the 
choice of the background model used to perform our measurement requires care.
In Sec.~\ref{s:likfits} we used the $p$-value of the sideband fits as a 
criterion to select the baseline background model, but the arbitrariness 
of that choice and the dependence of the background shape on the characteristics of the
data might affect the result. To estimate the resulting systematic uncertainties 
we perform fits to the experimental data using in turn each of the 286 shapes of background 
models described in section \ref{s:BGscan}. We then create a histogram of the results
obtained for the $b$-JES by each background shape, using the $p$-value
of their respective sideband fits as a weight.
We take the root-mean-square difference with respect to the most probable value, in the resulting distribution, as our estimate of 
the systematic uncertainty on the $b$-JES factor. This results in an absolute error of 
$^{+0.012}_{-0.006}$.

Two additional sources of systematic uncertainty are associated with
the background modeling. The first one is due to the finite
statistics of background templates used to derive the probability
density functions for our unbinned likelihood fit. To estimate the size of this effect
we perform different set of pseudo-experiments
using as input the background model which provides the best fit to experimental
data in the sideband region of the dijet mass distribution. 
For each set of pseudo-experiments we fluctuate with Poisson smearing the
number of events in each bin of the background distribution and measure the resulting 
bias on the fitted $b$-JES factor. The mean
systematic uncertainty on the $b$-JES is the found to be $\pm 0.011$.

A different potential source of systematic uncertainty comes from
correcting the background shape for signal contamination. As
described in section~\ref{s:BGscan}, we correct the background
templates for the presence of signal by subtracting the expected
signal assuming a data/MC $b$-JES of 1.00. We 
estimate the systematics related to that assumption by performing
again pseudo-experiments.
An additional $\pm 0.005$ systematic uncertainty is attributed to the $b$-JES factor
from the correction.

\subsection {Systematics related to Monte Carlo signal templates}\label{systbJES_MC}

For the pseudo-experiments described in Sec.~\ref{s:testproc}, signal
events used in the construction of pseudo-data samples are drawn
from the signal p.d.f.
That procedure is appropriate as long as we are testing the fitting procedure 
and estimating if the closeness of the signal to the background turn-on could yield 
any biases.
An additional source of uncertainty may be associated to the differences between the original
Monte Carlo $Z \to b\bar{b}$ dijet mass template and the signal
p.d.f used in the unbinned likelihood procedure. To study that effect, we
perform a further set of pseudo-experiments by drawing signal
events from the original MC template. 
We estimate a systematic uncertainty of $\pm 0.003$ on the $b$-JES factor from that source.

A second source of systematic uncertainty is related to the finite
statistics of the MC signal templates. To estimate this
error we use a similar procedure to that used for the modeling of
the background shape. Pseudo-experiments give a $\pm 0.002$ mean
systematic uncertainty on the $b$-JES.

\subsection { Other sources of uncertainty}\label{s:systbJES_other}

There are a few additional sources of systematic uncertainty in the determination 
of the $b$-jet energy scale which, however, are not included in our quoted total 
systematic error. In fact, the use of the measurement of a $b$-JES implies a choice 
of certain parameters and models in the simulation of Monte Carlo events: if the same choices 
are made as those we used in the generation of the $Z$ Monte Carlo sample 
(choice of generator, PDF set, and specific settings for initial and final state radiation 
modeling), there is no need to consider any systematic uncertainties affecting the $b$-JES 
due to those sources.

We nonetheless evaluate how a different choice of those
parameters changes the value of the $b$-JES. We use $Z \to b\bar{b}$ dijet
mass templates from samples of Monte Carlo with increased and
decreased initial state radiation (ISR) and final state radiation
(FSR) in pseudo-experiments to estimate the systematic
uncertainty on our measurement due to these sources. 
The result is an estimate of, respectively, $\pm 0.004$ and $\pm 0.012$ systematic uncertainty on the $b$-JES factor.

Additionally, we estimate the effect of the variation of the PDF set
on reconstructed $Z \to b\bar{b}$ dijet mass templates and thus on the
$b$-jet energy scale determination.
The estimate of the uncertainty is obtained by looking at the
difference between results obtained using the default CTEQ5L set of PDF 
and those obtained with the MRST72 set. Additionally, MRST72 and 
MRST75 sets derived using different $\Lambda_{QCD}$ values are compared, 
and 20 eigenvectors defining the CTEQ6M set are independently varied by $\pm 1$ 
standard deviation. Differences in pseudo-experiments resulting from these
different PDF's are added in quadrature to obtain a total
uncertainty of $\pm 0.005$ on the fitted $b$-JES factor.

\subsection {Final result on the $b$-jet energy scale}

To summarize, table~\ref{t:syst_uncert} details the sources of
systematic uncertainty on the $b$-jet energy scale measurement. All
sources are assumed to be uncorrelated; thus a total systematic
uncertainty is calculated as the sum in quadrature of the various
sources. Table~\ref{t:syst_other} shows other sources of
uncertainties which are not included in the $b$-JES factor
measurement.

Our final result for the $b$-jet energy scale factor is

\begin{center}
$k = 0.974 \pm 0.011(\mbox{stat}) ^{+0.017}_{-0.014}(\mbox{syst}) = 0.974 ^{+0.020}_{-0.018}(\mbox{total})$.
\end{center}

\noindent This $k$ factor refers to $R=0.7$ jets
corrected with level 5 standard CDF jet corrections. In addition, it has
been extracted from $b$-quark jets tagged by the SecVtx algorithm,
with transverse energy mostly in the range $22<E_T<50$ GeV, and it is relevant for
comparison between data collected by the CDF experiment from 2003 to
2005 and corresponding Pythia Monte Carlo simulations. 

The energy range relevant to $Z$ production is indeed restricted, though
not dramatically different from that of jets from top quark decay.
We believe our measurement of the $b$-jet energy scale can be used in top mass analyses, provided that 
the same jet cone definition is used ($R=0.7$ jets).
Analyses such as the top mass measurement in the dilepton channel could reduce sizably their systematic uncertainty by exploiting the correlations
between the standard JES, the $b$-jet energy scale and the measured top mass.
Unfortunately the $k$ factor we measured is not directly usable in analyses that use a different jet cone size ($R=0.4$ for example).
This is due to the fact that one of the largest systematic uncertainty in the generic jet energy scale correction is cone-size dependent (the level 7 jet correction).
Thus, we cannot constrain, for example, the $R=0.4$ jet cone uncertainties using a $k$ factor extracted from $R=0.7$ jets.
A future measurement of the $b$-jet energy scale using different jet cones could then be useful.

In the meantime, we believe our result can be used as
an independent constraint in measurements which are sensitive to
high $P_T$ $b$-jets and that use similar jet definitions as ours (jet $E_T$ range, cone size, etc.).



\section {Cross section determination}\label{s:xs}

In this section we evaluate the cross section for $Z$ boson production multiplied by 
the branching ratio of $Z$ boson decay to $b$-quark pairs using the number of
events of signal returned by a fit to the dijet mass distribution.

\subsection{Method of measurement}

The procedure to extract the signal is slightly different with respect to what is done for the $b$-jet energy scale measurement. First of all, since we aim to measure the $Z$ boson cross section we cannot use a constraint to the number of expected signal events in the likelihood fit. Second, we cannot use the $Z\to b\bar{b}$ cross section value to estimate, and subtract, the number of events of signal that fall into the kinematical regions used to construct the dijet mass background shape (see section \ref{s:effsig}). One simple way to solve this problem is to perform an iterative fitting procedure to correct the signal contamination in the background, without any prior knowledge of the $Z$ cross section. We use the following iterative method: an unbinned likelihood fit is performed to the selected data using an uncorrected background shape. In this fit the $b$-jet energy scale is let free and no constraint is applied to the expected number of events of signal. From the number of events of signal thus measured we can extrapolate (from $Z\to b\bar{b}$ Monte Carlo) the number of signal events in the background data samples, and subtract the resulting contribution. A new dijet invariant mass shape is then constructed from these background corrected samples. This procedure converges after a few iterations to a stable number of fitted signal events. To improve the precision of the measurement, the normalization of the background shape is constrained to the data sideband dijet mass region ($m_{jj}<60$ GeV/$c^{2}$ and $m_{jj}>120$ GeV/$c^{2}$), at each step of the iterative procedure. 

Apart from the fitting procedure, the general method of measurement is left unchanged with respect to the $b$-jet energy scale measurement. A large set of background dijet invariant mass shapes is constructed scanning the $\Delta \Phi$ and $E_t^3$ space, as described in section \ref{s:templates}. Each background shape is used to fit the selected data with the usual unbinned likelihood procedure, with the difference that the fit is performed now iteratively and that the background shape is constrained to the data sideband. We take as our measurement of the number of signal events the result obtained with the background shape that best fits the sideband (after iteration), while the results obtained with the other background models contribute to the estimate of the systematic uncertainty.

The measurement performed with the background shape that best fits the data sideband yields $N_{sig} = 6467 \pm 504$ (stat) events of signal and a $b$-jet energy scale factor of $0.976 \pm 0.010$ (stat), in agreement with the result in Sec.~\ref{s:systs}. From the fitted number of signal events we extract a cross section using the formula
\begin{eqnarray*}
\sigma_Z \times B(Z \rightarrow b\bar{b}) = \frac{N_{sig}}{\epsilon_{kin} \cdot \epsilon_{tag} \cdot L}
\end{eqnarray*}
where $\sigma_Z$ and $B(Z \rightarrow b\bar{b})$ are respectively the $Z$ boson 
cross section and the branching ratio of $Z$ decaying into a pair of $b$-quarks. 
The signal acceptance after all kinematical selections is given by the term $\epsilon_{kin}$, while
the $b$-tagging efficiency is given by the term $\epsilon_{tag}$. Finally $L$ is the 
total integrated luminosity for the dataset on which the measurement is performed.

\subsection{Evaluation of systematic uncertainties}

The effect of all main sources of systematic uncertainties affecting
the measured cross section is summarized in table
\ref{t:systxsec}. Below we discuss each contribution separately.

The systematic error on total signal efficiency, $\epsilon$, due to
standard CDF jet energy corrections, is estimated from $Z\to
b\bar{b}$ MC where the energy of measured jets is shifted by $\pm
1\sigma_{c}$ of the standard jet energy correction. The relative error
on the signal efficiency is then calculated as
$\frac{|\epsilon_{+1\sigma_{c}}-\epsilon_{-1\sigma_{c}}|}{2
\epsilon}$ and amounts to $(1.6 \pm 1.1)\%$.

The effect of increased or decreased final state radiation (FSR) on
signal efficiency is evaluated using $Z\to b\bar{b}$ MC samples
generated with different FSR tunings. The relative uncertainty on
efficiency due to this source is found to be $(2.9 \pm 1.1)\%$. The
systematic uncertainty related to initial state radiation was
derived from $Z \to \ell^{+}\ell^{-}$ data/MC comparison (where $\ell$ is an electron or a muon, see Sec.~\ref{s:SysUncISR}) 
and yields an additional $3.8\%$ contribution to the total error.
The relative error on signal efficiency due to different parton
distribution function parameterizations (see Sec.~\ref{s:systbJES_other}) 
is estimated to be $(7.3 \pm 3.0)\%$.
An additional source of systematic error is related to the MC generator
dependence. We evaluate it by comparing Pythia and Herwig~\cite{herwig} $Z\to b\bar{b}$
samples. The resulting uncertainty on signal efficiency is found to be $(2.2 \pm 5.4)\%$.
To be conservative we take the error on this measured value as our systematic.

The systematic affecting the trigger efficiency measurement
was also estimated (see Sec.~\ref{s:nsig}) and found to be $9.1\%$.

Several sources of systematic uncertainty related to the data-driven
background modeling and the fitting procedure were taken into account. 
The first one derives from the finite statistics in the templates used in the background
model construction, and was estimated performing pseudo-experiments
in a similar way as in Sec.~\ref{s:systbJES_BG}. The mean
uncertainty on the number of fitted signal events is $8.3\%$.
A second systematic uncertainty is related to the sideband criteria we applied to select the ``best'' background model used to perform the cross section measurement. As in Sec.~\ref{s:systbJES_BG} we create a histogram of the fitted number of events of signal obtained with each different
background shapes using the $p$-value of their respective sideband fits as a weight.
This fitting procedure, performed with the iterative method described previously,
yields an uncertainty on the number of fitted $Z \to b\bar{b}$
events of $^{+34.3\%}_{-15.0\%}$.

We also estimated the effect on the number of fitted signal events due to the constraint of the background shape to the dijet mass spectrum sideband. In fact an additional systematical error can arise from the uncertainty on the background normalization or from the small leakage of the signal in the sideband. We evaluate this uncertainty performing pseudo-experiments where the background normalization is shifted by $\pm 1 \sigma_{B}$ ($\sigma_{B}$ is the statistical uncertainty on the background normalization). The mean relative error on the fitted number of events of signal is measured to be $4.2\%$.

Finally, a systematic due to the iterative background correction procedure is estimated. The iterative method relies on the extrapolation of the number of events of signal in the background templates, given a fitted number of events of signal in the data. These background correction functions are estimated from Monte Carlo and yield a statistical uncertainty. The systematic error due to this source is estimated from fits to the data performed with smeared correction functions. The mean uncertainty on the number of signal events is estimated to be $1.6\%$.

Two additional sources of systematic uncertainty must be accounted for, the $b$-jet tagging data/MC scale factor uncertainty ($8.7\%$) and the error on the integrated luminosity measurement ($5.9\%$).

\subsection{Results}

\noindent Using the integrated luminosity, trigger and kinematic
efficiencies, and related systematic uncertainties described above, we derive:
\begin{eqnarray*}
\sigma_{Z} \times B(Z \rightarrow b\bar{b}) & = & 1578 \pm 123 (\mbox{stat}) \ ^{+624}_{-391} (\mbox{syst}) \ \mbox{pb} \\
                                   & = & 1578 \ ^{+636} _{-410}(\mbox{total})\ \mbox{pb}
\end{eqnarray*}

\noindent The measured cross section is higher but consistent within the uncertainties with 
the NLO theoretical calculation~\cite{nlozxs} combined with the measured $Z \to b
\bar{b}$ branching ratio~\cite{pdg}, which predicts
$\sigma_{Z} \times B(Z \rightarrow b \bar{b}) = 1129 \pm 22$ pb.

While this measurement has no appreciable impact on our knowledge of
the production and decay mechanism of the $Z$ boson in hadronic
collisions, it does fill a gap in the picture of measurements of
Standard Model production processes of vector bosons. If future
searches for new physics --especially the Higgs boson and
Supersymmetric particles-- at the Tevatron and at the LHC prove
successful, final states with $b$-quark jets will be very important
to study and measure. The current measurement and the described
methodology provide a normalization point for the measurement of the
production rate of new $b \bar{b}$ resonances, or for setting a
limit on their cross section.


\section{Conclusions and perspectives}\label{s:conclusions}

We have shown in this paper how a sizable sample of $Z$ boson decays to $b$-quark pairs
has been extracted from proton-antiproton collisions provided by the
Tevatron collider. 
We have also described in detail the method we used to obtain a precise
measurement of the $b$-jet energy scale from the shape of the dijet mass
distribution of the selected data.

The measurement can be used to reduce the dominant systematic
uncertainty in many of the CDF analyses which determine the top
quark mass, provided that they use a similar jet definition to ours.
The precise knowledge of the $b$-jet energy scale is of benefit also to
analyses attempting to reconstruct new resonances with decay to $b$-quark
jets, such as a low-mass Standard Model Higgs boson.

The signal has also been used to measure the cross section for $Z$ boson production using the $b \bar b$
final state: the result is $\sigma_Z \times B(b \bar b) =
1578^{+636}_{-410}$ pb.

\clearpage

\section*{Acknowledgments}

We thank the Fermilab staff and the technical staffs of the participating institutions 
for their vital contributions. This work was supported by the U.S. Department of Energy 
and National Science Foundation; the Italian Istituto Nazionale di Fisica Nucleare; the 
Ministry of Education, Culture, Sports, Science and Technology of Japan; the Natural 
Sciences and Engineering Research Council of Canada; the National Science Council of the 
Republic of China; the Swiss National Science Foundation; the A.P. Sloan Foundation;
 the Bundesministerium f\"ur Bildung und Forschung, Germany; the Korean Science and 
Engineering Foundation and the Korean Research Foundation; the Science and Technology 
Facilities Council and the Royal Society, UK; the Institut National de Physique 
Nucleaire et Physique des Particules/CNRS; the Russian Foundation for Basic Research; 
the Comisi\'on Interministerial de Ciencia y Tecnolog\'{\i}a, Spain; the European 
Community's Human Potential Programme; the Slovak R\&D Agency; and the Academy of Finland.
\vskip .4cm

The authors also wish to thank Lina Galtieri, Thomas Junk and Thomas Wright for 
their help in the preparation of the manuscript.


\clearpage


{\bf Figures captions}

Figure~\ref{f:btagdistr}:\\ {\em Invariant mass of the charged tracks in the vertex for
a sample of jets in single-tagged events (top) and for the tagged jet in double-tagged
events (bottom). The templates show the relative fraction of $b$,
$c$, and light quark or gluon jets estimated from the sample composition fit.}

Figure~\ref{f:etturnons_ets}:\\ {\em Jet $E_T$ distributions for the two leading jets in experimental
data. Left: corrected $E_T$ of the leading jet for all events (continuous line)
and events passing the preliminary selection (dashed line). Right: same, for the second
jet.}

Figure~\ref{f:kinvars}:\\ {\em Distributions (normalized to unity) of the variables used to optimize the kinematical selection, 
for data and Monte Carlo (Pythia). Left: azimuthal angle between the leading jets; right: corrected $E_T$ of the third jet.}

Figure~\ref{f:cZpt_060}:\\ {\em Left: $P_{T}$ distribution of the $Z$ boson for electron
data and MC. Right: corrected $E_{T}$ distribution of the leading jet for electron data and MC. All distributions are normalized to unity.}

Figure~\ref{f:cjetEta_cut}:\\ {\em Left: $\eta$ distribution of the leading jet for electron data and MC.
Right: $\Delta \phi$ distributions of the electron pair. All distributions are normalized to unity.}

Figure~\ref{f:BGPDFex}:\\ {\em Example of a data-driven background dijet mass distribution fit
with a Pearson IV function plus an error function.}

Figure~\ref{f:exsignalPDF}:\\ {\em Signal dijet mass distribution fitted with a sum of
three gaussian functions (for k=1.0).}

Figure~\ref{f:pexp1_SFOUT}:\\ {\em Mean fitted $b$-JES factor as a function of the input
scale factor.}

Figure~\ref{f:dijets_pp_FITRESULT_GC}:\\ {\em Results of the constrained unbinned likelihood fit
performed on double-tagged dijet data (points). A 
Gaussian constraint of $4630 \pm 681$ on the fitted number of signal events is
applied. The data-driven background shape and Monte Carlo signal
p.d.f are shown (hatched functions). The fit returns $5621
\pm 436$ signal events and a $b$-JES of $0.974 \pm 0.011$. The inset
on the upper right shows the data minus the background distribution
(points) and the signal shape normalized to the fitted
number of events of signal.}

\clearpage



\begin{table}[p]
\begin{center}
\begin{tabular}{|l|r|}
\hline
Selection level          & {Events}  \\
\hline
\hline
Total analyzed events     &  39 147 479 \\
\hline
Pass jet $E_T$ cuts       &  23 950 515      \\
Pass jet $|\eta_d|$ cuts  &  21 420 308      \\
Both jets taggable       &  18 128 488       \\
One tagged jet           &   6 205 578       \\
Two tagged jets        &     699 590       \\
Pass $\Delta\Phi_{12}$ and $E_t^3$ cuts & 267 246 \\
\hline
\end{tabular}
\caption{ \em Statistics of the analyzed data at different levels of selection.
          \label{t:stats_prelim}}
\end{center}
\end{table}


\begin{table}[p]
\centering
\begin{tabular}{|l|c|}
\hline
Source of uncertainty & Relative uncertainty \\
\hline
\hline
Statistical & 1.0\% \\
Luminosity  & 5.9\% \\
Data/MC $b$-tagging SF & 8.7\% \\
Z\_BB trigger simulation (calorimeter) & 1.8\% \\
Z\_BB trigger simulation (tracks)      & 8.9\% \\
ISR uncertainty                        & 3.8\% \\
Modeling of FSR                        & 2.9\% \\
\hline
Total uncertainty & 14.7\% \\
\hline
\end{tabular}
\caption{\em Uncertainties on the expected number of signal events.
  \label{t:SigAccUncertainty}}
\end{table}


\begin{table}[p]
\centering
\begin{tabular}{|l|c|}
\hline
Systematic source & $b$-JES factor \\
\hline
\hline
Background choice      & +0.012 -0.006 \\
Background statistics  & 0.011 \\
Background correction  & 0.005 \\
Monte Carlo template   & 0.003 \\
Monte Carlo statistics & 0.002 \\
\hline
Total              & +0.017 -0.014 \\
\hline
\end{tabular}
\caption{\em Summary of systematic uncertainties on the $b$-jet energy
scale. The total uncertainty is
obtained by adding the individual contributions in quadrature.
         \label{t:syst_uncert}}
\end{table}


\begin{table}[p]
\centering
\begin{tabular}{|l|c|}
\hline
Source & $b$-JES factor \\
\hline
\hline
Monte Carlo ISR        & 0.004 \\
Monte Carlo FSR        & 0.012 \\
Monte Carlo PDF's      & 0.005 \\
\hline
\end{tabular}
\caption{\em Other sources of uncertainties that are not included in the $b$-jet
energy scale measurement.
         \label{t:syst_other}}
\end{table}


\begin{table}[p]
\centering
\begin{tabular}{|l|l|c|}
\hline
Systematic source  & Method & Value \\
\hline
\hline
\multicolumn{3}{|c|}{Kinematical uncertainty} \\
\hline
JES          & $\pm 1\sigma_{c}$ jet corrections & 1.6\% \\
ISR          & $Z\rightarrow \ell^+\ell^-$ data/MC & 3.8\% \\
FSR          & MC                            & 2.9\% \\
PDF          & MC reweighting                & 7.3\% \\
MC gen.      & Pythia vs Herwig              & 5.4\% \\
Trigger      & Low $E_t$ QCD data            & 9.1\% \\
\hline
\multicolumn{2}{|l}{Relative uncertainty} & 13.8\% \\
\hline
\hline
\multicolumn{3}{|c|}{$b$-tagging efficiency} \\
\hline
Tagging eff. (two tags) & data/MC scale factor & 0.903 $\pm$ 0.079 \\
\hline
\multicolumn{2}{|l}{Relative uncertainty}   & 8.7\% \\
\hline
\hline
\multicolumn{3}{|c|}{Luminosity} \\
\hline
\multicolumn{2}{|l}{Total luminosity}       & 584.0 $\pm$ 34.5 \\
\multicolumn{2}{|l}{Relative uncertainty}   & 5.9\% \\
\hline
\hline
\multicolumn{3}{|c|}{Total signal acceptance uncertainty} \\
\hline
\multicolumn{2}{|l}{Relative uncertainty}  & 17.3\% \\
\hline
\hline
\multicolumn{3}{|c|}{Background systematics} \\
\hline
BG shape modeling & Pseudo-experiments & 8.3\% \\
BG normalization  & Pseudo-experiments & 4.2\%\\
BG model choice   & Sideband fit & +34.3\% -15.0\% \\
Iteration procedure & fit to data & 1.6\% \\
\hline
\end{tabular}
\caption{\em Summary of all sources of systematic uncertainties on $\sigma_{Z} \times B(Z \rightarrow b\bar{b})$ measurement.
  \label{t:systxsec}}
\end{table}


\begin{thebibliography}{99}
\bibitem{cdfsystem} The CDF reference frame uses cylindrical coordinates, where $\theta$ 
and $\phi$ are the polar and azimuthal angles with respect to the proton beam. The pseudorapidity
is defined as $\eta = -\ln \tan (\theta/2)$, the transverse energy $E_T = E \sin \theta$, and the
transverse momentum $P_T = P \sin \theta$.
\end{thebibliography}
\end{document}